\documentclass[11pt,letterpaper]{article}
\usepackage[T1]{fontenc}
\usepackage{lmodern}
\usepackage[margin=0.9in,headheight=15pt]{geometry}
\usepackage{amsmath,amssymb,amsthm,mathtools,booktabs,array}
\usepackage{enumitem,xcolor,fancyhdr,hyperref}
\usepackage{placeins}
\usepackage{natbib}
\usepackage{setspace}

\definecolor{inkblue}{HTML}{193A59}
\hypersetup{colorlinks=true,linkcolor=inkblue,urlcolor=inkblue,citecolor=inkblue,
pdftitle={Two-Way Fixed-Effects Panel Quantile Regression},
pdfsubject={Unsmoothed check loss, heterogeneous incidental biases, and detailed proofs at general quantiles}}
\setlist[enumerate]{leftmargin=*,itemsep=5pt,topsep=5pt}
\allowdisplaybreaks[2]

\newcommand{\E}{\mathbb E}\newcommand{\Pp}{\mathbb P}
\newcommand{\R}{\mathbb R}\newcommand{\ind}{\mathbf1}
\newcommand{\op}{o_p}\newcommand{\Op}{O_p}
\newcommand{\A}{\mathcal A}\newcommand{\D}{\mathcal D}
\newcommand{\M}{\mathcal M}\newcommand{\norm}[1]{\left\|#1\right\|}
\newcommand{\numberthis}{\addtocounter{equation}{1}\tag{\theequation}}

\newtheorem{theorem}{Theorem}
\newtheorem{lemma}{Lemma}

\theoremstyle{remark}\newtheorem{remark}{Remark}

\title{Two-Way Fixed-Effects Panel Quantile Regression\thanks{ We thank Iv\'an Fern\'andez-Val, Whitney Newey, and Pedro H. C. Sant'Anna for helpful comments. All errors are our own.}

}
\author{
	Harold D. Chiang\thanks{Harold D. Chiang: hdchiang@wisc.edu. Department of Economics, University of Wisconsin-Madison, William H. Sewell Social Science Building, 1180 Observatory Drive,	Madison, WI 53706, USA} 
	\quad 
	Antonio F. Galvao\thanks{	Antonio F. Galvao:  agalvao@msu.edu. 
Department of Economics,
Michigan State University, 
110 Marshall-Adams Hall, 
486 W. Circle Drive 
East Lansing, MI 48824, USA} 
	\quad 
Chia-Min Wei\thanks{Chia-Min Wei: cwei69@wisc.edu. Department of Economics, University of Wisconsin-Madison, William H. Sewell Social Science Building, 1180 Observatory Drive,	Madison, WI 53706, USA} 
    \quad 
Wei Biao Wu\thanks{Wei Biao Wu: wbwu@uchicago.edu. Department of Statistics, University of Chicago, George Herbert Jones Laboratory, 5747 South Ellis Avenue, Chicago, IL 60637, USA.}
}
\date{November 2, 2026}

\begin{document}

\maketitle\vspace{-18pt}
\begin{abstract}
This paper studies two-way fixed-effects quantile regression for panel data with individual and time
effects, estimated with the unsmoothed check loss. Individuals are independent and may be heterogeneous, and
no moments of the errors are required. When the numbers of individuals and periods grow at the same rate, the
slope estimator is asymptotically normal at rate $\sqrt{NT}$, with incidental biases of orders $T^{-1}$ and
$N^{-1}$ whose analytic form we derive. Unlike the one-way case, where each individual effect is fitted from that
individual's observations alone, the two sets of effects are fitted jointly and do not decouple. The proof
therefore approximates the additive fit by a weighted projection of the quantile scores, extracts the
heterogeneous bias from an individual-scaling identity obtained through linear-programming duality and
concavity, and then uses convexity to localise the slope. We propose a feasible analytic bias correction and a randomised
split-panel jackknife, with consistent variance estimators and valid inference.
\end{abstract}

\smallskip


\onehalfspacing

\newpage

\section{Introduction}\label{sec:intro}

Panel data sets have become widely available and very popular in empirical research. They provide a large number of data points, allow analysis of dynamics of adjustment, as well as control for individual-specific and time-specific heterogeneity. The literature on nonlinear panel data models with individual-specific fixed effects includes \citet{HahnNewey04}, \citet{FernandezVal05,FernandezVal09}, \citet{HonoreKyriazidou00}, \citet{DhaeneJochmans15} and \citet{HonoreWeidner25}. Nevertheless, the literature considering both individual-specific and time-specific effects is more restricted; see, e.g., \citet{FernandezValWeidner16}.

Yet additive models with two sets of fixed effects are a workhorse of empirical economics. A prominent example is the
model of \citet{AbowdKramarzMargolis99} for matched employer--employee data, in which the log wage of a worker is the
sum of a worker effect, an effect of the employing firm, a linear index of time-varying characteristics and an error.
Estimates of this model underlie much of the evidence on firm pay premiums, worker sorting and wage inequality
\citep[e.g.,][]{CardHeiningKline13}. Because differences between firm effects are identified only from workers who
move between firms, and most firms have few such movers, the estimated variance components suffer from a form of the
incidental parameter problem, known as limited mobility bias, which tends to overstate the variance of the firm
effects and to understate their correlation with the worker effects
\citep{AndrewsGillSchankUpward08,KlineSaggioSolvsten20,BonhommeEtAl23}.

Mean regressions, moreover, cannot show whether a firm's pay premium is the same throughout the wage distribution.
\citet{bonhomme2019distributional} model earnings distributions with two-sided worker--firm heterogeneity, grouping
firms into a few classes and treating worker heterogeneity as discrete. A quantile regression with two sets of fixed
effects leaves the effects unrestricted, but it faces an incidental parameter problem from each set of effects. Unlike limited mobility bias, which concerns quadratic functions of the estimated effects
in a linear model, this problem biases the estimator of the slope itself. The model studied here poses it in a simple
form, with individual and time effects in a complete design in which every individual is observed in every period, so
that each individual effect is estimated from $T$ observations and each time effect from $N$. In
matched employer--employee data, by contrast, most worker--firm pairs are never observed, and how accurately the
effects can be estimated depends on the network formed by workers' moves between firms \citep{JochmansWeidner19};
\citet{Verdier20} studies inference on the slope of linear models with such sparsely matched data. Extending our
results to such data is left for future work.

\citet{koenker} introduced panel data quantile regression (QR) models that identify and estimate conditional quantiles while controlling for individual-specific fixed effects (see also \citet{kato} and \citet{galvao}).
\citet{galvaokato16} analyse a smoothed quantile estimator for large panels and propose analytic and jackknife bias corrections, extending the methods for general nonlinear panels. The literature has expanded recently. \citet{cfw} propose a two-way quantile-effect procedure based on distribution regression. \citet{chiangwu} derive an unsmoothed one-way quantile expansion using a rank-score identity, and suggest bias correction methods. \citet{ArellanoHigginsWeidner26} study QR and instrumental variable QR in static panel data models with additive individual fixed effects.

Despite these advances, the statistical properties of the two-way fixed-effects QR estimator remain largely unexplored. Unlike linear mean regression, QR is based on a nonsmooth objective function, and the individual and time effects cannot generally be eliminated through differencing or within transformations. Moreover, estimating both sets of effects introduces two sources of incidental parameter bias, whose interaction must be accounted for in the asymptotic analysis. These difficulties are particularly important when individuals are heterogeneous, as is common in economic panel data, because the distributions of the regressors and errors, including their conditional densities at the quantile of interest, may differ across individuals. Existing results for one-way panel quantile regression \citep{chiangwu,ArellanoHigginsWeidner26} do not directly extend to the combined effects of these features.

This paper develops estimation and inference for two-way fixed-effects QR for panel data, with both individual-specific and time-specific effects, using the unsmoothed check loss at a fixed quantile level.
We develop an asymptotic theory for the unsmoothed two-way fixed-effects QR estimator considering large panels, i.e., where both the cross-sectional and time-series dimensions grow at the same rate. We consider a model with additive individual and time effects and a common slope parameter at a fixed quantile level. Our framework allows individuals to be independent but not identically distributed, accommodating heterogeneity in the joint distributions of regressors and errors. We establish a Bahadur linear representation for the estimator, and derive its limiting distribution. The estimator converges at the standard $\sqrt{NT}$ rate, but its asymptotic distribution contains two incidental parameter bias components, of orders $T^{-1}$ and $N^{-1}$, arising from the estimation of individual and time effects, respectively. Since these biases are of the same order as the sampling uncertainty, conventional inference that ignores them is generally invalid.

A central contribution of the paper is the explicit characterisation of these two bias components under cross-sectional heterogeneity. We show that the individual-effects bias depends on an average of individual-specific density-normalised derivatives, whereas the time-effects bias depends on the corresponding pooled density quantities. Consequently, the two bias coefficients generally differ, even though both arise from estimating additive fixed effects. Hence, the two-way conclusions here are not direct consequences of existing results for one-way panel QR, such as those of \citet{chiangwu} and \citet{ArellanoHigginsWeidner26}. The main argument derives the two heterogeneous bias contributions directly. A weighted quadratic approximation alone identifies the improvement in the additive fit. To extract a weighted score from that improvement, we scale each individual of the error array and use concavity of the improvement as a function of the scaling parameter. This supplies an explicit individual-weighted identity that a proof for identically distributed individuals would not need.
The convexity arguments follow the approach of \citet{hjortpollard} and \citet{knight}.

Building on the asymptotic expansion, we develop two approaches to bias correction and practical inference. First, we propose an analytic correction that estimates the leading bias components using kernel estimators of the conditional density quantities and their derivatives. To accommodate cross-sectional heterogeneity, we allow the bandwidths to vary across individuals and propose a nearest-neighbour procedure based on the fitted residuals. We establish consistency of the estimated bias coefficients as well as the estimated asymptotic covariance matrix, and show that the resulting bias-corrected estimator has a centred Gaussian limiting distribution. 

The second approach is a split-panel jackknife that removes the leading incidental parameter biases without explicitly estimating their coefficients. Because individuals need not be identically distributed, we introduce a randomised partition of the cross-sectional units to preserve the heterogeneous population composition of the sub-panels. We establish the asymptotic normality of the resulting jackknife estimator and develop a variance estimator constructed from the dual solutions of the full-panel and sub-panel quantile regressions. This procedure provides a feasible alternative for inference that accounts for the additional dispersion of the jackknife estimator in finite samples.


We investigate the finite-sample performance of the proposed procedures through Monte Carlo experiments that allow the distributions of regressors and errors to differ across individuals. The designs include Gaussian and asymmetric error distributions, heterogeneous scales, and settings in which the two incidental parameter bias components differ substantially. The simulation results provide numerical evidence supporting the theoretical characterisation of the bias as well as the importance of accounting for cross-sectional heterogeneity. The analytic correction substantially reduces the bias of the conventional estimator while maintaining similar dispersion, and the resulting confidence intervals exhibit coverage closer to their nominal levels. The split-panel jackknife also largely removes the bias, although its finite-sample variability can be larger than that of the original estimator. The proposed jackknife variance estimator improves the coverage of confidence intervals by accounting for this additional dispersion.

The rest of the paper proceeds as follows. Section~\ref{sec:model} states the model,
the estimator and the assumptions. Section~\ref{sec:main} states the slope
theorem and discusses the two bias terms. Section~\ref{sec:bias_correction} justifies an analytic bias correction
and a randomised split-panel jackknife with a variance estimator built from its
own fits, and describes the computation. Section~\ref{sec:simulations} reports simulations.
Appendix~\ref{sec:proofslope} proves the slope theorem. Appendix~\ref{sec:auxiliary} collects the deterministic
quadratic and duality results, the uniform rate of the fitted effects, and the rank of the design.
Appendices~\ref{sec:prooffeasible} to~\ref{sec:proofvariance} prove the results of Section~\ref{sec:bias_correction}.

\subsection*{Notation}

Let $\mathbb{R}$ denote the real line.
For a finite-dimensional vector $a$, let $\|a\|$ denote its Euclidean norm. For a matrix $A$, let $A'$ denote its transpose, $\|A\|$ its spectral norm, $\operatorname{tr}A$ its trace and $\lambda_{\min}(A)$ its smallest eigenvalue. For symmetric matrices, $A\preceq B$, or equivalently $B\succeq A$, means that $B-A$ is positive semidefinite, and $A\succ0$ that $A$ is positive definite. An array $v=(v_{it})$ indexed by $i\le N$ and $t\le T$ is treated as a vector in $\R^{NT}$, with Euclidean norm $\norm v_2$ and maximum norm $\norm v_\infty=\max_{i,t}|v_{it}|$.
We write $\ind(\cdot)$ for the indicator function, $\#A$ for the number of elements of a finite set $A$, $a\odot b$ for the elementwise product of two vectors, $a\vee b=\max(a,b)$ and $a\wedge b=\min(a,b)$. The law of $(x_{it},\varepsilon_{it})$ for individual $i$ is $P_i$ (Assumption A1 below), and $\E_i$, $\Pp_i$ and $\operatorname{Var}_i$ denote expectation, probability and variance under $P_i$.
Convergence in distribution is written $\Rightarrow$ and convergence in probability $\to_p$; $\mathcal N(\mu,\Sigma)$ is the normal law with mean $\mu$ and covariance matrix $\Sigma$. For random quantities, $X_{NT}=\op(r_{NT})$ means that $X_{NT}/r_{NT}\to_p0$, and $X_{NT}=\Op(r_{NT})$ that $X_{NT}/r_{NT}$ is bounded in probability. For positive sequences, $a_{NT}\asymp b_{NT}$ means that $a_{NT}/b_{NT}$ is bounded away from zero and infinity. Unless stated otherwise, $C$ and $c$ denote generic positive constants that may depend on $\tau$ and on the constants in the assumptions, but not on $i$, $t$, $N$ or $T$, and whose values may change from one use to the next. The probability spaces
we consider are assumed to be standard Borel.
Throughout the remainder of the paper, all asymptotic statements refer to regimes in which both $N$ and $T$ diverge to infinity. 

\section{Model, estimator, and assumptions}\label{sec:model}

\subsection{Model and estimator}\label{ssec:estimator}

Let $\{(y_{it},x_{it}')' : i = 1,\ldots,N; t = 1,\ldots,T\}$ be panel data. We consider a panel QR model with individual and time effects
\begin{equation}
\label{eq:model0}
Q_{\tau}(y_{it}\mid x_{it}) = \alpha_{i0} + \gamma_{t0} + x_{it}'\beta_{0},
\end{equation}
where $\tau \in (0,1)$ is a quantile of interest,  $y_{it}$ is a dependent variable, $x_{it}$ is a $p$-dimensional vector of explanatory variables,\footnote{The vector $x_{it}$ has no constant component. A constant would be absorbed by the individual effects and would make the matrix $H_N$ in \eqref{eq:pooled} singular, which Assumption A3 rules out. A grand intercept can be separated from the individual effects by constraining these to sum to zero as well; see the paragraph on a constant regressor below.}$\alpha_{i0}$ is a scalar individual-specific effect for each $i$, $\gamma_{t0}$ is a scalar time-specific effect for each $t$, and $Q_{\tau}(y_{it}\mid x_{it})$ is the conditional $\tau$-quantile of $y_{it}$ given $x_{it}$. 
In general, each of $\alpha_{i0}$, $\gamma_{t0}$ and $\beta_{0}$ can depend on $\tau$, but we assume $\tau$ to be fixed throughout the paper and suppress such a dependence for notational simplicity. The model is semiparametric in the sense that the functional form of the conditional distribution of $y_{it}$ given $x_{it}$ is left unspecified and no parametric assumption is made on the relation between $\alpha_{i0}$, $\gamma_{t0}$ and $x_{it}$. The number of individuals is denoted by $N$ and the number of time periods is denoted by $T = T_{N}$, which may depend on $N$. In what follows, we omit the subscript $N$ of $T_{N}$. It should be noted that the quantile is a nonlinear functional of the random variable, so in general the conventional differencing out strategy is not appropriate for the QR case.  

The model in equation \eqref{eq:model0} can be rewritten as the following regression
\begin{equation}
\label{eq:model}
y_{it}=\alpha_{i0}+\gamma_{t0}+x_{it}'\beta_{0}+\varepsilon_{it},
\qquad \sum_{t=1}^T\gamma_{t0}=0,
\end{equation}
where $Q_{\tau}(\varepsilon_{it}\mid x_{it})=0$. The normalisation $\sum_{t=1}^T\gamma_{t0}=0$ is needed because the individual and time effects are identified only up to a common shift: replacing $\alpha_{i0}$ by $\alpha_{i0}+c$ and $\gamma_{t0}$ by $\gamma_{t0}-c$ leaves every conditional quantile in \eqref{eq:model0} unchanged. Under this normalisation, $\gamma_{t0}$ is the deviation of period $t$ from the average time effect over the $T$ periods of the panel, and $\alpha_{i0}$ includes that average. The estimator in \eqref{eq:estimator} below imposes the same constraint. Another normalisation, such as $\gamma_{10}=0$, would leave the slope estimator unchanged but would change the meaning of the estimated effects. Since the normalisation involves all $T$ periods, the effects form triangular arrays that may change with $T$, a dependence that we suppress. Otherwise, individual and time effects may be arbitrary deterministic arrays.

We consider the fixed-effects estimation of $\beta_{0}$, which is implemented by treating individual and time effects as parameters to be estimated. 
Then the standard two-way fixed-effects QR (TWFEQR) estimation solves 
\begin{equation}
\label{eq:estimator}
(\widehat\alpha,\widehat\gamma,\widehat\beta)
\in\operatorname*{arg\,min}_{\alpha\in\mathbb{R}^{N},\gamma\in\mathbb{R}^{T},\beta\in\mathbb{R}^{p},\sum_{t=1}^{T}\gamma_{t}=0}
\sum_{i=1}^{N}\sum_{t=1}^{T}\rho_\tau(y_{it}-\alpha_{i}-\gamma_{t}-x_{it}'\beta),
\end{equation}
where $\rho_\tau(u)=u\{\tau-\ind(u\le0)\}$ is the check function of \citet{koenkerbassett}, and $\psi_\tau(u)=\tau-\ind(u\le0)$.
For compact notation, write $\rho=\rho_\tau$ and $\psi=\psi_\tau$ below. The subdifferential at
zero is $\partial\rho(0)=[\tau-1,\tau]$.
The objective function in \eqref{eq:estimator} is convex and piecewise linear, so a minimiser exists by the linear-programming formulation of quantile regression (\citet{koenkerbook}). It need not be unique, and we select one minimiser as follows.

\emph{Selection rule.} The minimiser is chosen measurably and equivariantly under permutations of the individuals, permutations of the times, and shifts of the responses by additive arrays $r_i+s_t$. One such rule, which we call the least-squares (LS) rule, chooses, among all minimising fitted arrays $(\alpha_i+\gamma_t+x_{it}'\beta)_{i,t}$, the one with the least sum of squared residuals. Since the set of minimising fitted arrays is compact and convex and the squared loss is strictly convex, the LS rule selects a unique fitted array. This array determines $(\widehat\alpha,\widehat\gamma,\widehat\beta)$ uniquely when the regressors have full rank modulo the effects, an event whose probability tends to one; a measurable and equivariant convention on the complementary event, such as the slope of least norm among those giving the selected fitted array, does not affect the results (Appendix~\ref{ssec:rank}). The selection rule restricts the estimator, not the data, and it is not an identification condition: identification rests on the normalisation in \eqref{eq:model} and on Assumptions A2 and A3 below. By equivariance, the selected slope does not change when an additive array $r_i+s_t$ is added to the responses, so the true effects may be taken to be zero in the proofs and in the simulations of Section~\ref{sec:simulations}.

\emph{A constant regressor and double centring.}\label{ssec:constantreg} Suppose instead that the regressor is
$(1,w_{it}')'$ with coefficient $(\theta_0,b')'$, and both effect sums are constrained to zero, so that $\theta_0$ is
a grand intercept. Denote the centred individual effects by $\alpha_{i}^{\mathrm{dc}}$. The corresponding
single-centring intercepts are
\begin{equation}
\eta_i=\alpha_{i}^{\mathrm{dc}}+\theta_0,\qquad
\theta_0=\bar\eta,\qquad
\alpha_{i}^{\mathrm{dc}}=\eta_i-\bar\eta.
\label{eq:normalization}
\end{equation}
This reparameterisation is a bijection between the two normalised parameter spaces and preserves every
residual. The fit of $b$ is therefore that of \eqref{eq:estimator} with regressors $w_{it}$, and the results below
apply to the nonconstant coefficient $b$. The estimated grand intercept, the average of the $N$ fitted
$\eta_i$, is an average of fitted effects; a $\sqrt{NT}$ law for it requires a finer analysis of that average
than the uniform rate of Lemma~\ref{lem:uniformrate}, and is not studied here.

\subsection{Assumptions}\label{ssec:assumptions}

This section provides conditions under which the TWFEQR estimator described above has a limiting normal distribution with a bias in the mean when $N$ and $T$ diverge to infinity and $N/T \to \kappa$. The conditions are, to some extent, standard in the literature.  We will also characterise the bias, and Sections~\ref{sec:feasible} and~\ref{sec:spj} propose bias correction estimation and practical inference procedures.

\begin{enumerate}[label=\textbf{A\arabic*.}]
\item \textbf{Independent heterogeneous individuals, iid time observations.}
For each fixed $i$, $\{(x_{it},\varepsilon_{it}):t\ge1\}$ is independent and identically distributed (iid) with law $P_i$, and these sequences are independent across $i$. The regressors $x_{it}\in\R^p$ have fixed dimension $p$.

\item \textbf{Uniform conditional $\tau$-quantile and density conditions.}
For every $i$, $\Pp_i(\varepsilon\le0\mid x)=\tau$ and the conditional
law has a density $f_i(u\mid x)$. Write $F_i(u\mid x)=\Pp_i(\varepsilon\le u\mid x)$
for its conditional distribution function, chosen measurable in $x$ for each $u$. For fixed $r_0,c_0,C_0>0$, uniformly
in $i$ and almost surely in $x$,
\[
f_i(0\mid x)\ge c_0,\qquad
\sup_{|u|\le r_0}\{|f_i(u\mid x)|+|f_i'(u\mid x)|\}\le C_0.
\]
The density is continuously differentiable on this interval and, for a
common modulus $\omega_0(\delta)\downarrow0$,
\[
|f_i'(u\mid x)-f_i'(v\mid x)|\le\omega_0(|u-v|)
\quad (|u|,|v|\le r_0).
\]

\item \textbf{Bounded regressors and limiting moments.}
$\norm x\le K$ almost surely uniformly in $i$. 
Define the following
\begin{gather}
f_{i0}(x)=f_i(0\mid x),\quad f_{i1}(x)=f_i'(0\mid x),\quad
a_i=\E_i f_{i0}(x),\quad
\mu_i=\frac{\E_i[f_{i0}(x)x]}{a_i},\quad z_{it}=x_{it}-\mu_i,
\label{eq:rowmoments}\\
H_i=\E_i[f_{i0}(x)(x-\mu_i)(x-\mu_i)'],\quad
V_i=\E_i[(x-\mu_i)(x-\mu_i)'],\quad
c_i=\E_i[f_{i1}(x)(x-\mu_i)],
\label{eq:hvc}\\
H_{N}=\frac{1}{N}\sum_{i=1}^{N}H_{i},\quad V_{N}=\frac{1}{N}\sum_{i=1}^{N}V_{i},\quad
\bar a_{N}=\frac{1}{N}\sum_{i=1}^{N} a_i,\quad
\bar c_{N}=\frac{1}{N}\sum_{i=1}^{N} c_i,
\label{eq:pooled}\\
B_N=-\frac{\tau(1-\tau)}{2}H_N^{-1}\frac{1}{N}\sum_{i=1}^{N}\frac{c_i}{a_i^2},\qquad
D_N=-\frac{\tau(1-\tau)}{2}H_N^{-1}\frac{\bar c_N}{\bar a_N^2}.
\label{eq:bias}
\end{gather}
The limits $H := \lim_{N\to\infty}H_{N}$ and $V := \lim_{N\to\infty}V_{N}$ exist and are nonsingular, and
$\bar a_N\to\bar a>0$. Put $\Omega=\tau(1-\tau) H^{-1}VH^{-1}$.
For an uncentred limiting normal law, additionally
assume the limits $B := \lim_{N\to\infty}B_{N}$ and $D := \lim_{N\to\infty}D_{N}$ exist. These last two limits are unnecessary for the finite-$N$ bias-centred conclusions.

\item \textbf{Rates.} $N,T\to\infty$ with $N/T\to\kappa\in(0,\infty)$.
\end{enumerate}

Assumptions A1--A4 adapt conditions that are standard in the literature to the two-way model. Condition A1 is the two-way analogue of condition (A1) of \citet{kato}, which they note is the same as Condition~1(i) of \citet{FernandezVal05}. It is stated for $(x_{it},\varepsilon_{it})$ rather than for $(y_{it},x_{it})$ because $y_{it}$ contains the time effect $\gamma_{t0}$: the pairs $(y_{it},x_{it})$ are independent over $t$, but they are identically distributed only when all time effects are equal. In condition A1 we exclude temporal dependence to focus on the simplest case and to highlight the difficulties arising from the TWFEQR estimator.
Assumption A1 does not require identical distributions over $i$. This distinction changes the bias formula, and it is the reason why the split-panel jackknife of Section~\ref{sec:spj} splits the individuals at random.

Conditions in Assumption A2 impose restrictions on conditional densities, and are standard in the QR literature. We do not impose a global identification condition such as condition (A3) of \citet{kato}, which bounds from below, uniformly in $i$, the increase in each individual's expected check loss when $(\alpha_i,\beta)$ moves a fixed distance away from $(\alpha_{i0},\beta_{0})$. Here identification comes from A2 and A3: the positive conditional densities at zero and the nonsingular limit $H$ make the expected objective locally strictly convex at the true parameters, and the convexity of the objective turns this local curvature into the global bounds used in the proof of Theorem~\ref{thm:slope} (Appendix~\ref{sec:proofslope}). Because $H_N$ is an average over individuals, A3 does not require the regressors of every individual to vary over time, whereas condition (A3) of Kato, Galvao, and Montes-Rojas fails for an individual whose regressors are constant over time.

The laws $P_i$ may depend on $N$: the results below hold for arrays $P_i=P_{i,N}$ that satisfy A2 and $\norm x\le K$ with common constants and the limits in A3, since their proofs use only these bounds and limits and make
distributional statements at each $(N,T)$. Such arrays arise, for example, when the number of distributional types grows with $N$, and in design SH of Section~\ref{sec:simulations}, in which the law of individual $i$ depends on $(i-1)/(N-1)$.

In Assumption A3, $B_N$ is the coefficient of the bias that the estimation of the individual effects induces in the slope, and $D_N$ that of the time effects. Their minus signs correspond to the positive
regression sign in \eqref{eq:model}. The constants $a_i$ denote marginal densities at zero, not individual intercepts. Under A4, $N+T\asymp\min(N,T)\asymp\sqrt{NT}$. All constants in the proof can be chosen uniformly over individuals, but may depend on the fixed $\tau$. No moment condition on the errors is used. A quantile process, uniform coverage over a continuum of levels, and sequences approaching zero
or one require additional arguments and are not asserted here. All quantities $a_i,\mu_i,H_N,V_N,B_N,D_N$ and $\Omega$ are evaluated at the fixed level $\tau$; we omit the dependence for simplicity.

\section{Main asymptotic result}\label{sec:main}

In this section, we turn our attention to the limiting statistical properties of the TWFEQR estimator in equation \eqref{eq:estimator}.

The next result presents a linear Bahadur representation and the limiting distribution of the TWFEQR estimator when $N$ and $T$ grow at the same rate. 

\begin{theorem}\label{thm:slope}
Under A1--A4, we have that
\begin{equation}
\widehat\beta-\beta_{0}
=H_N^{-1}\frac1{NT}\sum_{i,t}z_{it}\psi(\varepsilon_{it})
+\frac{B_N}{T}+\frac{D_N}{N}
+\op((NT)^{-1/2}).
\label{eq:slope}
\end{equation}
Consequently,
\begin{equation}
\sqrt{NT}\left(\widehat\beta-\beta_{0}-\frac{B_N}{T}-\frac{D_N}{N}\right)
\Rightarrow\mathcal N(0,\Omega).
\label{eq:slopeclt}
\end{equation}
If $B_N\to B$ and $D_N\to D$, the uncentred limiting mean is
$\sqrt\kappa B+\kappa^{-1/2}D$.
\end{theorem}

The proof, in Appendix~\ref{sec:proofslope}, has four steps. First, with the slope at $\beta_0$ or within
$O((NT)^{-1/2})$ of it, the fitted individual and time effects are uniformly of order $\min(N,T)^{-1/3}$ and close to
an explicit weighted projection of the quantile scores, which also gives the improvement in the check loss from
fitting the effects (Lemmas~\ref{lem:additive} and~\ref{lem:quadraticapprox}). Second, linear-programming duality
writes the profiled criterion as a maximum over a fixed polytope; rescaling the errors of each individual then makes
the weighted score sum that the expansion needs a supergradient of the improvement in fit, and evaluating that
improvement yields the two bias coefficients (Lemma~\ref{lem:scaling}). Third, the subgradient of the profiled
criterion splits into the leading score, a term that is centred given the errors and negligible, and the bias, at
$\beta_0$ and at nearby slopes. Fourth, convexity of the profiled criterion turns this local expansion into
\eqref{eq:slope} for every minimiser, and the Lindeberg--Feller central limit theorem gives \eqref{eq:slopeclt}.

Theorem~\ref{thm:slope} implies $\widehat\beta-\beta_0=\Op((NT)^{-1/2})$, so the slope estimator is consistent. Lemma~\ref{lem:uniformrate} in Appendix~\ref{sec:auxiliary} adds that the fitted effects are uniformly consistent; indeed,
$\max_i|\widehat\alpha_i-\alpha_{i0}|+\max_t|\widehat\gamma_t-\gamma_{t0}|=\Op(\min(N,T)^{-1/3})$. A separate
consistency result is therefore not needed, and the proof does not use one as a preliminary step: convexity of the
profiled criterion places every minimiser within $\Op((NT)^{-1/2})$ of $\beta_0$ directly
(Appendix~\ref{sssec:convexity}). This is also why no global identification condition such as condition (A3) of
\citet{kato}, on which their consistency argument rests, is required here (Section~\ref{ssec:assumptions}).

Equation~\eqref{eq:slopeclt} centres $\widehat\beta-\beta_{0}$ at two bias terms, $B_N/T$ and $D_N/N$, defined in equation \eqref{eq:bias} above. The bias term $B_N/T$ arises
from estimating the $N$ individual effects, each from $T$ observations. In the same fashion, the bias $D_N/N$ is due to estimating the $T$ time effects, each from $N$ observations. The fitted effects enter the bias corrections of Sections~\ref{sec:feasible} and~\ref{sec:spj} only through the
residuals, for which the uniform rate of Lemma~\ref{lem:uniformrate} in Appendix~\ref{sec:auxiliary} suffices.

To gain some intuition into the bias expressions, we examine the bias formulas in two simple cases: identically distributed individuals, and two distributional types of individuals. 

First, consider the case of identically distributed individuals. If $P_i=P$ for all $i$, then $a_i=a$, $c_i=c$, $H_N=H$, and
$V_N=V$. Therefore, the formulas in \eqref{eq:bias} reduce to
\begin{equation*}
B_N=D_N=-\frac{\tau(1-\tau)}{2a^2}H^{-1}c.
\end{equation*}
Consequently, the slope bias is the common coefficient times $T^{-1}+N^{-1}$,
under the sign convention in \eqref{eq:model}.

Next, consider a case where the two heterogeneous bias terms differ. Suppose we have two distributional types with limiting proportions $\lambda$ and $1-\lambda$. Then, the bias formulas  in \eqref{eq:bias} are
\begin{align*}
B&=-\tfrac{\tau(1-\tau)}{2}H^{-1}
\left\{\lambda c_A/a_A^2+(1-\lambda)c_B/a_B^2\right\},\\
D&=-\tfrac{\tau(1-\tau)}{2}H^{-1}
\frac{\lambda c_A+(1-\lambda)c_B}{\{\lambda a_A+(1-\lambda)a_B\}^2},
\end{align*}
where $a_A,c_A,H_A$ and $a_B,c_B,H_B$ are the moments in \eqref{eq:rowmoments}--\eqref{eq:hvc} of the two types, and $H=\lambda H_A+(1-\lambda)H_B$.
The individual bias averages the inverse-square densities before aggregation, and the time bias uses the inverse square of the pooled density. Even when
$\lambda c_A+(1-\lambda)c_B=0$, the individual bias can remain nonzero if $a_A\ne a_B$. Therefore, simply replacing the homogeneous density by its pooled
average does not provide the correct heterogeneous extension.

Moreover, note that if $c_i=0$ for every individual, both biases vanish. Sufficient conditions
are $f_i'(0\mid x)=0$ almost surely for every individual, or independence of $x$ and $\varepsilon$ within each individual. At the median, $\tau=1/2$, conditional symmetry about zero implies the first condition. For continuous conditional distributions, symmetry about zero is compatible with the maintained quantile normalisation only at the median. Independence of $x$ and $\varepsilon$ within each individual allows different, asymmetric error distributions across individuals. Then, $\mu_i=\E_i[x]$ and $f_{i1}$ is constant within the individual, so $c_i=0$. The bias coefficients refer to the centring of the limit law in Theorem~\ref{thm:slope}, not to the expectation of the estimator, which the theory does not compute.

\section{Bias correction and inference}\label{sec:bias_correction}

As usual in the literature, an important issue with the limiting distribution of $\sqrt{NT}(\widehat\beta-\beta_{0})$ not being centred at zero is that standard confidence intervals based on the asymptotic approximation will be incorrect. The limiting mean $\sqrt\kappa B+\kappa^{-1/2}D$ in Theorem~\ref{thm:slope} is of the same order as the standard deviation; its first term grows with $N/T$ and its second with $T/N$, so even small coefficients $B$ and $D$ matter when the panel is far from square. In this section, we suggest two bias correction methods to the TWFEQR estimator together with corresponding practical inference procedures.

\subsection{Feasible analytic bias correction}\label{sec:feasible}

First, we suggest an analytic bias correction. The conditions stated allow for consistent estimation of the bias coefficients and covariance matrix. For a fixed quantile $\tau$, define the fitted residuals
\begin{equation*}
\widehat\varepsilon_{it}=y_{it}-x_{it}'\widehat\beta
-\widehat\alpha_{i}-\widehat\gamma_{t} .
\end{equation*}
Next, we wish to estimate the quantities described in equations \eqref{eq:rowmoments}--\eqref{eq:bias}. To do so, let $K_0$ be a nonnegative, symmetric, twice continuously differentiable kernel of integral one that vanishes outside $[-1,1]$. For example,
\[
K_0(u)=\frac{35}{32}(1-u^2)^3\ind\{|u|\le1\}
\]
has the needed properties.

Each individual is allowed to have its own bandwidths: $h_i$ for its density-weighted moments and $g_i$ for its density derivative. For a fraction $\pi\in(0,1]$ let $\widehat h_i(\pi)$ be the $\lceil\pi T\rceil$-th smallest of
$|\widehat\varepsilon_{i1}|,\ldots,|\widehat\varepsilon_{iT}|$, so that the window
$|\widehat\varepsilon_{it}|\le\widehat h_i(\pi)$ holds $\lceil\pi T\rceil$ of the $T$ observations of individual
$i$ (more only if absolute residuals are tied), and take
\[
h_i=\widehat h_i(\pi_a),\qquad g_i=\widehat h_i(\pi_c),\qquad
\pi_a=T^{-1/5},\qquad \pi_c=T^{-1/7}.
\]
These nearest-neighbour bandwidths follow the scale of each individual's errors. The fractions $\pi_a$ and $\pi_c$
carry no units and have the usual orders for estimating a density and its derivative from $T$ observations.
With these bandwidths the estimates below are equivariant in the scale of the residuals, i.e., if every residual is
multiplied by a positive constant, then the estimated bias coefficients, $\widehat{B}$ and $\widehat{D}$, are multiplied by that constant and the estimated covariance matrix, $\widehat\Omega$, by its square. No floor on the density estimates is needed.
More generally, the theorem below holds for any bandwidths, possibly depending on the data, that satisfy the
following condition.

\begin{enumerate}[label=\textbf{A\arabic*.},resume]
\item \textbf{Bandwidths.} For constants
$0<\underline c\le\bar c<\infty$ and positive deterministic sequences $\pi_a$ and $\pi_c$, the bandwidths satisfy
\begin{equation}
\begin{gathered}
\Pp\bigl\{\underline c\,\pi_a\le h_i\le\bar c\,\pi_a\ \text{and}\
\underline c\,\pi_c\le g_i\le\bar c\,\pi_c\ \text{for all }i\bigr\}\to1,\\
\pi_a\to0,\qquad \pi_c\to0,\qquad \min(N,T)^{1/3}\pi_a\to\infty,\qquad
\min(N,T)^{1/3}\pi_c^2\to\infty .
\end{gathered}
\label{eq:bandwidth}
\end{equation}
\end{enumerate}

Note that a deterministic bandwidth common to all individuals, $h_i=g_i=h>0$ with $h\to0$ and
$\min(N,T)^{1/3}h^2\to\infty$, is a special case of Assumption A5, with $\pi_a=\pi_c=h$ and $\underline c=\bar c=1$. 

The terms $a_{i}$, $\mu_{i}$, $z_{it}$, $c_{i}$, $H$, and $V$ can be estimated by
\begin{align}
\widehat a_i&=\frac1{Th_i}\sum_{t=1}^{T}K_0(\widehat\varepsilon_{it}/h_i),
&\widehat\mu_i&=\frac{\sum_{t=1}^{T} x_{it}K_0(\widehat\varepsilon_{it}/h_i)}
{\sum_{t=1}^{T}K_0(\widehat\varepsilon_{it}/h_i)},
\label{eq:ahat}\\
\widehat z_{it}&=x_{it}-\widehat\mu_i,
&\widehat c_i&=-\frac1{Tg_i^2}\sum_{t=1}^{T}
\widehat z_{it}K_0'(\widehat\varepsilon_{it}/g_i),
\label{eq:chat}\\
\widehat H&=\frac{1}{N}\sum_{i=1}^{N}\frac1{Th_i}\sum_{t}K_0(\widehat\varepsilon_{it}/h_i)
\widehat z_{it}\widehat z_{it}',
&\widehat V&=\frac1{NT}\sum_{i,t}\widehat z_{it}\widehat z_{it}'.
\label{eq:hvhat}
\end{align}
The minus sign in \eqref{eq:chat} is required for estimation of a density
derivative at zero. 

Let $\widehat{\bar a}=N^{-1}\sum_{i=1}^{N}\widehat a_i$ and $\widehat{\bar c}=N^{-1}\sum_{i=1}^{N}\widehat c_i$. 
The bias coefficients and the covariance matrix are estimated by
\begin{gather}
\widehat B=-\frac{\tau(1-\tau)}{2}\widehat H^{-1}
\frac{1}{N}\sum_{i=1}^{N}\frac{\widehat c_i}{\widehat a_i^2},\qquad
\widehat D=-\frac{\tau(1-\tau)}{2}\widehat H^{-1}
\frac{\widehat{\bar c}}{(\widehat{\bar a})^2},\qquad
\widehat\Omega=\tau(1-\tau)\widehat H^{-1}\widehat V\widehat H^{-1}.
\label{eq:biasestimators}
\end{gather}
The definitions may be assigned arbitrary values on the vanishing-probability event that some bandwidth or some $\widehat a_i$ is zero or $\widehat H$ is not invertible.

Finally, we define the analytic bias-corrected TWFEQR estimator, $\widehat\beta_{\mathrm{ABC}}$, by
\begin{gather}
\widehat\beta_{\mathrm{ABC}}
=\widehat\beta-\widehat B/T-\widehat D/N.
\label{eq:abc}
\end{gather}

The next result establishes feasible inference under the analytic bias correction.

\begin{theorem}\label{thm:feasible}
Under A1--A4, the bandwidths $h_i=\widehat h_i(\pi_a)$ and $g_i=\widehat h_i(\pi_c)$ satisfy A5,
with $\underline c=1/(4C_0)$ and $\bar c=2/c_0$, for all fractions $\pi_a$ and $\pi_c$ that satisfy the second line in \eqref{eq:bandwidth};
these include $\pi_a=T^{-1/5}$ and $\pi_c=T^{-1/7}$. Under A1--A5,
\begin{equation}
\max_i\left\{ |\widehat a_i-a_i|+
\norm{\widehat\mu_i-\mu_i}+\norm{\widehat c_i-c_i}\right\}=\op(1),
\label{eq:uniformdensityconsistency}
\end{equation}
and $\widehat H-H_N\to_p0$, $\widehat V-V_N\to_p0$,
$\widehat B-B_N\to_p0$, $\widehat D-D_N\to_p0$,
$\widehat\Omega\to_p\Omega$. In addition,
\begin{equation}
\sqrt{NT}(\widehat\beta_{\mathrm{ABC}}-\beta_{0})
\Rightarrow\mathcal N(0,\Omega).
\label{eq:abcclt}
\end{equation}
\end{theorem}

Given the results in Theorem~\ref{thm:feasible}, practical inference is straightforward. In particular, for each
fixed $\ell\neq0$, the studentised statistic $\ell'(\widehat\beta_{\mathrm{ABC}}-\beta_0)/\{\ell'\widehat\Omega\ell/(NT)\}^{1/2}$
converges in distribution to $\mathcal N(0,1)$, because $\ell'\Omega\ell>0$ (Appendix~\ref{sec:prooffeasible}), so
that $\ell'\widehat\beta_{\mathrm{ABC}}\pm1.96\{\ell'\widehat\Omega\ell/(NT)\}^{1/2}$ is an asymptotic 95 per cent
confidence interval for $\ell'\beta_0$.

Computing the analytic bias-corrected estimator and its covariance matrix requires bandwidths.
Theorem~\ref{thm:feasible} fixes only their rates, through the fractions $\pi_a$ and $\pi_c$, and leaves the
constants free; $T^{-1/5}$ and $T^{-1/7}$ are the usual orders for estimating a density and its derivative from $T$
observations. An optimal choice of the constants for the bias and variance estimates, for instance along the lines of
\citet{HallHorowitz90}, is outside the scope of this paper, and we leave it for future research.
Section~\ref{sec:simulations} reports how the results change when either fraction is multiplied by $2/3$ or $3/2$,
and that a deterministic bandwidth common to all individuals, which Assumption~A5 also allows, performs poorly in the
designs studied.

\subsection{A split-panel jackknife that preserves heterogeneity}\label{sec:spj}

Next, we consider the split-panel jackknife of \citet{DhaeneJochmans15}, originally proposed for panels with
individual fixed effects, in its half-panel form. It removes the bias of the TWFEQR estimator $\widehat\beta$ without
estimating the bias coefficients.

Split the periods into $\mathcal T_1=\{1,\ldots,T_1\}$ and $\mathcal T_2=\{T_1+1,\ldots,T\}$, where $T_2=T-T_1$ and
$T_1,T_2\asymp T$, for instance $T_1=\lfloor T/2\rfloor$. Independently of the data, choose a uniform random
partition of the individuals into two sets $\mathcal R_1$ and $\mathcal R_2$ of sizes $N_1,N_2\asymp N$. We also
write $\mathcal T_j$ for the sub-panel of all individuals in the periods of $\mathcal T_j$, and $\mathcal R_j$ for the
sub-panel of all periods of the individuals in $\mathcal R_j$. For each of these four sub-panels $\mathcal P$, let
$\widehat\beta^{\mathcal P}$ be the slope estimate obtained by applying \eqref{eq:estimator} to the observations of
$\mathcal P$, with its own individual and time effects, the latter summing to zero over its periods. With the
size-weighted averages
\[
\bar\beta_T=\frac{T_1}{T}\widehat\beta^{\mathcal T_1}+\frac{T_2}{T}\widehat\beta^{\mathcal T_2},\qquad
\bar\beta_N=\frac{N_1}{N}\widehat\beta^{\mathcal R_1}+\frac{N_2}{N}\widehat\beta^{\mathcal R_2},
\]
define the split-panel jackknife TWFEQR estimator as
\begin{equation}
\widehat\beta_{\mathrm{SPJ}}
=3\widehat\beta-\bar\beta_T-\bar\beta_N.
\label{eq:spj}
\end{equation}

The following result summarises the limiting properties of the estimator in equation \eqref{eq:spj}.
\begin{theorem}[Randomised cross-sectional splitting]\label{thm:spj}
Under A1--A4, for the independently randomised individual split just defined, we have that
\begin{equation}
\sqrt{NT}(\widehat\beta_{\mathrm{SPJ}}-\beta_{0})
\Rightarrow\mathcal N(0,\Omega).
\label{eq:spjclt}
\end{equation}
Under A5, the estimator $\widehat\Omega$ in \eqref{eq:biasestimators} consistently
estimates $\Omega$, its asymptotic covariance matrix, as well (Theorem~\ref{thm:feasible}).
\end{theorem}

The jackknife removes both bias terms because each half panel has, to first order, the population moments of the
full panel, so that $\bar\beta_T$ has bias $2B_N/T+D_N/N$ and $\bar\beta_N$ has bias $B_N/T+2D_N/N$, up to terms of
smaller order (Appendix~\ref{sec:proofspj}). A deterministic division into the first and last halves of the individuals need not
preserve the heterogeneous population moments, and the cancellation is therefore not justified for that split
without an additional balance condition. Random allocation, or a deterministic split verified to balance the relevant moments, supplies the needed condition.
Temporal splitting requires no such randomisation here because each individual's observations are iid over time.

Next, we consider the variance estimator for the jackknife. Note that the asymptotic covariance matrix of the TWFEQR estimator in equation \eqref{eq:slopeclt} of Theorem~\ref{thm:slope} is the same as that in equation \eqref{eq:spjclt} in Theorem~\ref{thm:spj}.
The estimator $\widehat\Omega$ in equation \eqref{eq:biasestimators} estimates the first-order covariance of $\widehat\beta_{\mathrm{SPJ}}$, but in finite samples the jackknife is more dispersed than $\widehat\beta$, as the simulations of
Section~\ref{sec:simulations} show and as \citet{galvaokato16} also report for the jackknife of their one-way
smoothed estimator. We suggest an alternative that combines the dual solutions of the five fits in~\eqref{eq:spj}.

Write $\mathcal F$ for the full panel, so that the five fits in \eqref{eq:spj} are those to
$\mathcal P\in\{\mathcal F,\mathcal T_1,\mathcal T_2,\mathcal R_1,\mathcal R_2\}$. Also, let $N_{\mathcal P}$ and $T_{\mathcal P}$ be the numbers of individuals and periods of $\mathcal P$, and write $\mathcal T(t)$ and $\mathcal R(i)$ for the time half containing period $t$ and the individual half containing individual $i$.
Let 
\begin{equation*}
\widehat\varepsilon^{\,\mathcal P}_{it}=y_{it}-x_{it}'\widehat\beta^{\mathcal P}-\widehat\alpha^{\mathcal P}_i-\widehat\gamma^{\mathcal P}_t
\end{equation*}
be the residuals of the selected fit to $\mathcal P$.
For each fit, let $\widehat q^{\,\mathcal P}$ be a measurable choice of solution of the dual of its linear
programme, which maximises $\sum_{(i,t)\in\mathcal P}y_{it}q_{it}$ over arrays on the cells of $\mathcal P$ with zero
individual and time sums, $\sum_{(i,t)\in\mathcal P}x_{it}q_{it}=0$ and $\tau-1\le q_{it}\le\tau$. Such a choice exists by Lemma~\ref{lem:quantileduality}, applied to the span of the effects and the regressors. Among the finitely many partitions of the cells into those at $\tau-1$, those at $\tau$ and the rest, the conditions that a partition defines a unique feasible point and that this point maximises the objective are
Borel in the data, and the maximum is attained at a vertex, so the first qualifying partition gives a Borel choice. At the nonzero
residuals, $\widehat q^{\,\mathcal P}_{it}=\tau-\ind(\widehat\varepsilon^{\,\mathcal P}_{it}<0)$ (see the proof of Theorem~\ref{thm:spjvariance}
in Appendix~\ref{sec:proofvariance}); at the zero residuals it takes values in $[\tau-1,\tau]$ that satisfy the constraints, which determine them
at a nondegenerate vertex. With $\widehat z_{it}=x_{it}-\widehat\mu_i$ and $\widehat H$ the full-panel estimates
of Section~\ref{sec:feasible}, define the following estimator of $\Omega/(NT)$, the first-order covariance matrix
of $\widehat\beta_{\mathrm{SPJ}}$:
\begin{equation*}
\widehat{V}_J=\frac1{(NT)^2}\sum_{i,t}\widehat\phi_{it}\widehat\phi_{it}' , \qquad   
\widehat\phi_{it}=\widehat H^{-1}\widehat z_{it}\bigl(3\widehat q^{\,\mathcal F}_{it}
-\widehat q^{\,\mathcal T(t)}_{it}-\widehat q^{\,\mathcal R(i)}_{it}\bigr)
.
\end{equation*}
The coefficients follow from equation~\eqref{eq:spj}. A cell of a half enters that half's fit with weight
$(N_{\mathcal P}T_{\mathcal P})^{-1}$, and the half enters $\bar\beta_T$ or $\bar\beta_N$ with weight $N_{\mathcal P}T_{\mathcal P}/(NT)$, so that
the cell's term has coefficient $-1$ relative to $(NT)^{-1}$, for halves of any sizes, against $3$ for the full panel. Since $x_{it}-\widehat z_{it}$ depends only on the individual, and each $\widehat q^{\,\mathcal P}$ has zero individual sums and is
orthogonal to the regressors, the $\widehat\phi_{it}$ sum to zero, so that $\widehat V_J$ needs no centring. The
five fits are those of the jackknife itself; if the five dual solutions were equal, $\widehat V_J$ would be the
influence-function version of $\widehat\Omega/(NT)$.\footnote{Note that kernel estimates computed separately in each half, with its own bandwidth, would add smoothing-bias differences of larger order than the effect that $\widehat V_J$ is meant to capture; the full-panel estimates avoid them.}
The next result establishes the consistency of $\widehat V_J$ and the resulting size control of tests based on
$\widehat\beta_{\mathrm{SPJ}}$.

\begin{theorem}[Jackknife variance and size control]\label{thm:spjvariance}
Under A1--A5, for the independently randomised individual split of
Theorem~\ref{thm:spj},
\[
NT\,\widehat V_J\to_p\Omega,\qquad
\widehat V_J^{-1/2}(\widehat\beta_{\mathrm{SPJ}}-\beta_{0})\Rightarrow\mathcal N(0,I_p).
\]
Hence, for each significance level $\varpi\in(0,1)$, the null rejection probabilities of the Wald test of $\beta_{0}=b^\dagger$, which rejects when $(\widehat\beta_{\mathrm{SPJ}}-b^\dagger)'\widehat V_J^{-1}(\widehat\beta_{\mathrm{SPJ}}-b^\dagger)$
exceeds the $1-\varpi$ quantile of $\chi^2_p$, and, for each fixed $\ell\ne0$, of the test of
$\ell'\beta_{0}=r^\dagger$, which rejects when
$|\ell'\widehat\beta_{\mathrm{SPJ}}-r^\dagger|>\Phi^{-1}(1-\varpi/2)(\ell'\widehat V_J\ell)^{1/2}$ with $\Phi$ the
standard normal distribution function, converge to $\varpi$. If the limits in A3 are replaced by
$V_N\succeq v_0I_p$ for a fixed $v_0>0$, the convergence is uniform over the effect arrays and over the arrays of individual
laws, possibly changing with $N$, that satisfy A2 and $\norm x\le K$ with the same $r_0$, $c_0$, $C_0$, $\omega_0$ and $K$; that is, both tests have asymptotic size $\varpi$ over this class.
\end{theorem}

The class in the uniform statement keeps A1, A4, the selection rule of Section~\ref{ssec:estimator} and~A5, whose first line is to hold uniformly over the class, as it does for
the nearest-neighbour bandwidths (by the first claim of Theorem~\ref{thm:feasible} along every sequence in the
class, through the subsequence argument of its proof in Appendix~\ref{sec:proofvariance}; its constants $1/(4C_0)$ and $2/c_0$ are those of the
class); the uniform statement is at each
fixed $\beta_{0}$ satisfying the null; the confidence sets obtained by inverting the tests have asymptotic coverage
$1-\varpi$ in the same sense. The conclusions hold for any measurable choice among the dual solutions, and for any
estimates with $\max_i\norm{\widehat\mu_i-\mu_i}\to_p0$ and $\widehat H-H_N\to_p0$ (for the uniform statement,
along every sequence of effect arrays and individual laws in the class, that is, uniformly over it; the estimates of
Section~\ref{sec:feasible} qualify by Theorem~\ref{thm:feasible} and the argument of the proof). The definitions may be assigned
arbitrary values on the vanishing-probability event that $\widehat H$ or $\widehat V_J$ is not invertible. The
uniformity step is the subsequence argument for asymptotic size of \citet{acg}.

\begin{remark}[First and second order]\label{rem:spjvariance}
Conditionally on the split, the leading term of $\widehat\beta_{\mathrm{SPJ}}-\beta_{0}$, before the replacement of the
individual halves' inverses in the proof of Theorem~\ref{thm:spj}, is
$(NT)^{-1}\sum_{i,t}(2H_N^{-1}-H_{\mathcal R(i)}^{-1})z_{it}\psi(\varepsilon_{it})$, where
$H_{\mathcal R_j}=N_j^{-1}\sum_{i\in\mathcal R_j}H_i$; the time halves contribute $H_N^{-1}$ because their
population moments are those of the full panel. Its conditional covariance given the split, multiplied by $NT$, is
\[
\Omega_J=\frac{\tau(1-\tau)}{N}\sum_{i=1}^{N}(2H_N^{-1}-H_{\mathcal R(i)}^{-1})V_i(2H_N^{-1}-H_{\mathcal R(i)}^{-1}),
\]
and $\Omega_J-\tau(1-\tau) H_N^{-1}V_NH_N^{-1}=\Op(N^{-1})$ over the split. Indeed, with
$G_j=H_{\mathcal R_j}-H_N=\Op(N^{-1/2})$, $E_j=H_{\mathcal R_j}^{-1}-H_N^{-1}$ and
$V_{(j)}=N_j^{-1}\sum_{i\in\mathcal R_j}V_i=V_N+\Op(N^{-1/2})$, the terms of $\Omega_J$ linear in the $E_j$ are
exactly $-\tau(1-\tau)\sum_j(N_j/N)(E_jV_{(j)}H_N^{-1}+H_N^{-1}V_{(j)}E_j)$, the quadratic terms
$\tau(1-\tau)\sum_j(N_j/N)E_jV_{(j)}E_j$ are $\Op(N^{-1})$, and the identity
$(H+G)^{-1}=H^{-1}-H^{-1}GH^{-1}+H^{-1}GH^{-1}G(H+G)^{-1}$ with $\sum_j(N_j/N)G_j=0$ gives
$\sum_j(N_j/N)E_j=\Op(N^{-1})$, which, as $E_j(V_{(j)}-V_N)=\Op(N^{-1})$, makes the linear terms
$\Op(N^{-1})$ too. Differences between the individual halves' matrices, which change the first-order
covariance of a jackknife when the halves differ systematically, therefore play no role under A1 with a random
split, and $\widehat V_J$ and $\widehat\Omega$ estimate the same first-order covariance. They differ through the
dual values, which change between the fits at the cells at which a half-panel fit changes the sign of a residual or
has a zero residual; the share of these cells tends to zero by the proof of Theorem~\ref{thm:spjvariance}. Heuristically, these are the cells
that make the jackknife more dispersed than $\widehat\beta$: at a fixed slope, the second line of the score
decomposition~\eqref{eq:scoredecomp} is, given the errors, a sum of independent cell terms with mean zero, supported
where $q\neq\psi$, so its conditional variance is a sum of squares of the kind that enters $\widehat V_J$; the
covariances between cells that a fitted slope and fitted effects add are not estimated. This is not a proof.
Theorem~\ref{thm:spjvariance} is a first-order result, and the smoothing bias of the kernel estimates exceeds
the second-order differences in question, so neither $\widehat V_J$ nor $\widehat\Omega$ is accurate to that
precision; what can be checked is the ratio of the two, and Section~\ref{sec:simulations} reports it. Being
first-order, Theorem~\ref{thm:spjvariance} does not distinguish the two estimators: its size statement also holds
with $\widehat\Omega/(NT)$ in place of $\widehat V_J$, by Theorems~\ref{thm:feasible} and~\ref{thm:spj} and, for the
uniform part, the argument of its proof. Size control in this sense is asymptotic: it bounds no rejection
probability at a given $(N,T)$, where the two estimators differ through the second-order terms above, and
Section~\ref{sec:simulations} reports the coverage of both intervals, computed with the kernel estimates of Section~\ref{sec:feasible},
in the designs studied.
\end{remark}

\subsection{Computation}\label{sec:practice}

We finish this section with a note on computation of the TWFEQR estimator. For a specified $\tau$, equation \eqref{eq:estimator} is a linear programme, and in practice it is solved as a quantile
regression of $y_{it}$ on $x_{it}$, $N$ individual dummies, and $T-1$ time dummies, that is, with one period's
dummy dropped, which imposes $\gamma_1=0$ in place of $\sum_{t=1}^T\gamma_t=0$. The two normalisations give the
same sets of minimising fitted arrays and slopes, so the slope estimate is the same when the minimiser is unique; the effects differ by a common constant, and subtracting
the average of the fitted time effects from each of them, and adding it to each fitted individual effect, gives
the effects normalised as in \eqref{eq:model}. The design is sparse, and the simulations of
Section~\ref{sec:simulations} use the sparse interior-point algorithm of \citet{portnoykoenker} and
\citet{koenkerng} in the \textsf{R} package \texttt{quantreg} \citep{quantreg}. Theorem~\ref{thm:slope} holds for
every minimiser, so the selection rule of Section~\ref{ssec:estimator} matters only for how the effects are
reported. The kernel estimates of Section~\ref{sec:feasible} use the residuals of the full-panel fit, and the
half-panel fits of Section~\ref{sec:spj} apply the same estimator to the sub-panels. All intervals are pointwise in $\tau$; simultaneous coverage across several levels is
not proved here. The stochastic conclusions require no moments of the error distribution.

\section{Monte Carlo simulations}\label{sec:simulations}

This section examines the finite-sample performance of the procedures of Sections~\ref{sec:main},~\ref{sec:feasible} and~\ref{sec:spj} in designs in which the error laws, the regressor distributions and the densities at the quantile differ across individuals. Two findings stand out. First, the bias of the slope estimator follows $B_N/T+D_N/N$ of Theorem~\ref{thm:slope}, with its two distinct coefficients, at the central quantile levels in every design and at every level in a Gaussian design, whereas a formula that treats the individuals as identical is rejected. Second, the analytic correction removes about half of the bias without adding dispersion, and its intervals are close to nominal at $N=T=100$; the split-panel jackknife removes the first-order bias, but its intervals need the variance estimator of Theorem~\ref{thm:spjvariance}. The Supplemental Material (\texttt{twfe\_v4\_simstudy.pdf}) reports every configuration in Tables~S.I to~S.XII.

\subsection{Designs and implementation}\label{ssec:mcdesigns}

\emph{Designs.} The response is $y_{it}=x_{it}\beta_{0}+\varepsilon_{it}$ with $\beta_{0}=1$, $x_{it}\sim U(0,b_i)$ and
\begin{equation}
\varepsilon_{it}=s_i\,k_i\,(1+x_{it})\,(e_{it}-q_i),\label{eq:mcslope}
\end{equation}
where $e_{it}$ is drawn from a standard family, $s_i\in\{-1,1\}$ reflects it, $k_i>0$ is a scale and $q_i$ is the $\tau$-quantile of $e_{it}$ (its $(1-\tau)$-quantile when $s_i=-1$), so that $\Pp(\varepsilon_{it}\le0\mid x_{it})=\tau$ and the density at the quantile varies with the regressor. In design SN, $e_{it}\sim N(0,1)$, $s_i=1$, $b_i=2$, and $k_i=1$ for the first half of the individuals and $k_i=2$ for the second. In design SC, $e_{it}\sim\chi^2_3$ and $b_i=2$; the first half of the individuals have $(s_i,k_i)=(1,1)$ and the second half $(-1,2)$, so that their error laws are skewed in opposite directions. Design SC2 is SC with one fifth of the individuals of the first type, so that at the median $D_N=0$ but $B_N\ne0$, the case of Section~\ref{sec:main}. In design SH no two individuals share a law: $k_i=1+2u_i$ and $b_i=3-2u_i$ with $u_i=(i-1)/(N-1)$, and the $\chi^2_3$ errors are skewed to the right for odd $i$ and to the left for even $i$. SN is run at $\tau\in\{0.1,0.25,0.5\}$, SC and SH at $\tau\in\{0.1,0.25,0.5,0.75,0.9\}$, and SC2 at the median, which gives $14$ configurations. In all of them the asymptotic bias of $\widehat\beta$ at $N=T$, computed by quadrature from~\eqref{eq:rowmoments}--\eqref{eq:bias}, is at most $0.73$ asymptotic standard deviations in absolute value.

\emph{Implementation.} The panel sizes are $N=T\in\{50,100\}$, with $2000$ replications each, so that the Monte Carlo standard error of a coverage rate near $95$ per cent is $0.5$ percentage points. The individual and time effects are set to zero, which is no loss of generality under an equivariant selection rule (Section~\ref{ssec:estimator}). The check-loss problem is solved by the sparse interior-point algorithm of the \textsf{R} package \texttt{quantreg} (\citet{quantreg}; \citet{portnoykoenker}; \citet{koenkerng}; \citet{rcore}). The analytic correction $\widehat\beta_{\mathrm{ABC}}=\widehat\beta-\widehat B/T-\widehat D/N$ and the covariance estimator $\widehat\Omega$ use the triweight kernel and the nearest-neighbour (NN) bandwidths of Section~\ref{sec:feasible} with $\pi_a=T^{-1/5}$ and $\pi_c=T^{-1/7}$, fixed before the main runs from pilot runs on a subset of the same draws. The split-panel jackknife $\widehat\beta_{\mathrm{SPJ}}$ of Section~\ref{sec:spj} uses time halves and a random split of the individuals, and its variance estimator is $\widehat V_J$ of Theorem~\ref{thm:spjvariance}; the \emph{oracle} correction $\widehat\beta-B_N/T-D_N/N$ uses the true coefficients. All intervals have nominal level $95$ per cent, biases are reported on the $\sqrt{NT}$ scale, and MAD denotes the mean absolute deviation of the coverage from $95$ per cent over the configurations, in percentage points.

\subsection{Results}\label{ssec:mcslope}

Tables~\ref{tab:mcslope50} and~\ref{tab:mcslope100} report the results at $N=T=50$ and $100$. At $\tau\in\{0.25,0.5,0.75\}$ the bias of $\widehat\beta$ agrees with Theorem~\ref{thm:slope} in every design run at that level at $N=T=100$, where $|z|$ is at most $2.3$. In the Gaussian design SN the formula fits at every level and size ($|z|\le1.6$). The pooled prediction $\sqrt{NT}D_N(1/T+1/N)$, obtained by treating the individuals as identical, is rejected in each skewed design at some level; in SC2 at the median, where it predicts no bias, its $z$ is $4.8$ at $N=T=50$ and $3.2$ at $N=T=100$, against $1.7$ and $0.1$ for Theorem~\ref{thm:slope}. In the tails of the skewed designs there are sizable terms of higher order: at $\tau\in\{0.1,0.9\}$ in SC and SH the bias is $1.6$ to $2.3$ times the prediction at $N=T=50$ and $1.4$ to $1.8$ times at $100$.

\begin{table}[t]\centering
\caption{The slope at $N=T=50$: bias against Theorem~\ref{thm:slope}, corrections and coverage}\label{tab:mcslope50}
\small\setlength{\tabcolsep}{4.2pt}
\begin{tabular}{llrrrrrrrrrr}
\toprule
 & & \multicolumn{7}{c}{bias ($\times\sqrt{NT}$)} & \multicolumn{3}{c}{coverage (per cent)} \\
\cmidrule(lr){3-9}\cmidrule(lr){10-12}
Design & $\tau$ & Thm~\ref{thm:slope} & FE & (s.e.) & $z$ & $z_{\rm pooled}$ & ABC & SPJ & FE & ABC & SPJ$_J$ \\
\midrule
SN & 0.10 & 5.60 & 5.49 & (0.19) & $-$0.6 & $-$2.1 & 3.99 & $-$1.20 & 82.3 & 86.1 & 87.3 \\
SN & 0.25 & 1.87 & 1.70 & (0.14) & $-$1.2 & $-$1.9 & 0.86 & $-$0.39 & 91.5 & 92.6 & 92.0 \\
SN & 0.50 & 0.00 & $-$0.20 & (0.13) & $-$1.6 & $-$1.6 & $-$0.21 & $-$0.25 & 92.2 & 91.8 & 92.4 \\
\addlinespace
SC & 0.10 & 3.19 & 7.30 & (0.24) & 17.2 & 21.7 & 4.72 & $-$4.27 & 88.0 & 92.2 & 92.8 \\
SC & 0.25 & 0.94 & 2.49 & (0.24) & 6.4 & 11.9 & 0.30 & $-$1.40 & 91.5 & 92.0 & 92.1 \\
SC & 0.50 & $-$0.96 & $-$0.59 & (0.28) & 1.3 & 4.8 & $-$0.78 & $-$0.61 & 91.4 & 90.6 & 90.7 \\
SC & 0.75 & $-$2.68 & $-$3.45 & (0.29) & $-$2.6 & $-$3.5 & $-$1.11 & 0.18 & 92.1 & 92.3 & 92.0 \\
SC & 0.90 & $-$5.27 & $-$8.28 & (0.33) & $-$9.0 & $-$13.5 & $-$4.94 & 4.96 & 88.1 & 91.1 & 91.2 \\
\addlinespace
SC2 & 0.50 & 1.08 & 1.68 & (0.35) & 1.7 & 4.8 & 1.07 & 0.14 & 92.2 & 91.5 & 91.7 \\
\addlinespace
SH & 0.10 & 4.93 & 10.59 & (0.33) & 17.0 & 20.5 & 7.09 & $-$4.55 & 87.3 & 90.9 & 90.8 \\
SH & 0.25 & 1.79 & 2.83 & (0.32) & 3.3 & 6.9 & 0.10 & $-$2.74 & 92.2 & 92.1 & 91.6 \\
SH & 0.50 & $-$0.11 & $-$0.02 & (0.33) & 0.2 & 0.4 & 0.00 & $-$0.10 & 91.1 & 90.6 & 90.8 \\
SH & 0.75 & $-$1.98 & $-$2.70 & (0.33) & $-$2.2 & $-$5.7 & 0.11 & 2.98 & 92.3 & 92.1 & 92.7 \\
SH & 0.90 & $-$5.15 & $-$11.33 & (0.35) & $-$17.8 & $-$21.6 & $-$7.73 & 4.45 & 86.0 & 90.2 & 92.0 \\
\bottomrule
\end{tabular}

\par\medskip\begin{minipage}{0.95\linewidth}\footnotesize
\emph{Note}: Thm~\ref{thm:slope}: $\sqrt{NT}(B_N/T+D_N/N)$. Bias: Monte Carlo mean of $\sqrt{NT}(\widehat\beta_\bullet-\beta_{0})$ for $\widehat\beta$ (FE), with its standard error, and for $\widehat\beta_{\mathrm{ABC}}$ (ABC) and $\widehat\beta_{\mathrm{SPJ}}$ (SPJ); $z$: (FE $-$ Thm~\ref{thm:slope})/s.e.; $z_{\rm pooled}$: the same with the pooled prediction $\sqrt{NT}D_N(1/T+1/N)$. Coverage of $\widehat\beta_\bullet\pm1.96\{\widehat\Omega/(NT)\}^{1/2}$ (FE, ABC) and of $\widehat\beta_{\mathrm{SPJ}}\pm1.96\widehat V_J^{1/2}$ (SPJ$_J$). $2000$ replications. Tables~S.I to~S.III of the Supplemental Material report every configuration.
\end{minipage}
\end{table}

\begin{table}[t]\centering
\caption{The slope at $N=T=100$: bias against Theorem~\ref{thm:slope}, corrections and coverage}\label{tab:mcslope100}
\small\setlength{\tabcolsep}{4.2pt}
\begin{tabular}{llrrrrrrrrrr}
\toprule
 & & \multicolumn{7}{c}{bias ($\times\sqrt{NT}$)} & \multicolumn{3}{c}{coverage (per cent)} \\
\cmidrule(lr){3-9}\cmidrule(lr){10-12}
Design & $\tau$ & Thm~\ref{thm:slope} & FE & (s.e.) & $z$ & $z_{\rm pooled}$ & ABC & SPJ & FE & ABC & SPJ$_J$ \\
\midrule
SN & 0.10 & 5.60 & 5.64 & (0.18) & 0.2 & $-$1.4 & 3.57 & $-$0.36 & 85.8 & 90.3 & 91.2 \\
SN & 0.25 & 1.87 & 1.93 & (0.14) & 0.4 & $-$0.3 & 0.86 & $-$0.04 & 92.6 & 92.8 & 92.6 \\
SN & 0.50 & 0.00 & $-$0.11 & (0.13) & $-$0.8 & $-$0.8 & $-$0.11 & $-$0.10 & 93.4 & 93.1 & 93.2 \\
\addlinespace
SC & 0.10 & 3.19 & 5.82 & (0.21) & 12.7 & 17.9 & 2.84 & $-$2.81 & 91.8 & 95.3 & 94.9 \\
SC & 0.25 & 0.94 & 1.31 & (0.23) & 1.6 & 7.2 & $-$1.25 & $-$1.65 & 93.9 & 93.6 & 93.7 \\
SC & 0.50 & $-$0.96 & $-$0.58 & (0.27) & 1.4 & 5.0 & $-$0.44 & 0.24 & 93.7 & 93.2 & 94.0 \\
SC & 0.75 & $-$2.68 & $-$2.85 & (0.29) & $-$0.6 & $-$1.4 & 0.25 & 0.55 & 93.4 & 93.6 & 93.0 \\
SC & 0.90 & $-$5.27 & $-$7.20 & (0.30) & $-$6.4 & $-$11.3 & $-$2.86 & 2.22 & 92.1 & 94.9 & 95.6 \\
\addlinespace
SC2 & 0.50 & 1.08 & 1.11 & (0.34) & 0.1 & 3.2 & 0.26 & $-$0.55 & 92.7 & 92.3 & 92.8 \\
\addlinespace
SH & 0.10 & 5.00 & 8.82 & (0.30) & 12.6 & 16.6 & 4.58 & $-$2.92 & 90.6 & 93.6 & 95.1 \\
SH & 0.25 & 1.84 & 2.06 & (0.31) & 0.7 & 4.4 & $-$1.44 & $-$2.13 & 94.0 & 93.9 & 94.1 \\
SH & 0.50 & $-$0.05 & $-$0.30 & (0.33) & $-$0.8 & $-$0.7 & $-$0.28 & $-$0.61 & 93.9 & 93.7 & 93.8 \\
SH & 0.75 & $-$1.94 & $-$2.70 & (0.32) & $-$2.3 & $-$6.0 & 0.78 & 1.36 & 92.1 & 93.0 & 92.8 \\
SH & 0.90 & $-$5.11 & $-$8.76 & (0.31) & $-$11.9 & $-$16.1 & $-$4.44 & 3.54 & 90.6 & 94.1 & 94.2 \\
\bottomrule
\end{tabular}

\par\medskip\begin{minipage}{0.95\linewidth}\footnotesize
\emph{Note}: As in Table~\ref{tab:mcslope50}, at $N=T=100$ with $2000$ replications.
\end{minipage}
\end{table}

The analytic correction removes about half of this bias and adds no dispersion. Averaged over the $14$ configurations, the absolute bias in units of the standard deviation of $\widehat\beta$ falls from $0.33$ and $0.31$ at $N=T=50$ and $100$ to $0.19$ and $0.15$, against $0.13$ and $0.08$ for the oracle correction, while the standard deviation of $\widehat\beta_{\mathrm{ABC}}$ is $0.99$ to $1.02$ times that of $\widehat\beta$. The standard error is close to the asymptotic one, with $(\E\widehat\Omega/\Omega)^{1/2}$ between $0.94$ and $1.18$ at $N=T=100$, and the corrected intervals cover $86.1$ to $92.6$ and $90.3$ to $95.3$ per cent at the two sizes (MAD $3.9$ and $1.7$), against $82.3$ to $92.3$ and $85.8$ to $94.0$ per cent around $\widehat\beta$. The gain is in the tails, where the bias is largest. At $N=T=50$ all corrected intervals undercover, but so do the oracle intervals ($90.3$ to $93.3$ per cent): at that size the first-order variance $\Omega$ itself understates the variance of $\widehat\beta$. The estimated coefficients are less accurate than the corrected estimator suggests. In SN they are too small, mostly because of smoothing, and in SC at $\tau=0.25$, $\widehat D$ has the wrong sign (Table~S.IV). Multiplying either NN fraction by $2/3$ or $3/2$ gives MAD between $1.3$ and $6.6$; the smallest, at both sizes, is from multiplying $\pi_a$ by $3/2$ ($1.7$ and $1.3$, against $3.9$ and $1.7$ for the fractions used here; Tables~S.V to~S.VII). A deterministic bandwidth common to all individuals, which Assumption~A5 also allows, fails in these designs: with $h=\min(N,T)^{-1/12}$, between $0.68$ and $0.72$ here, every density estimate is at the floor $h$ that this rule places on them, $(\E\widehat\Omega/\Omega)^{1/2}$ is between $0.35$ and $0.70$, and the intervals cover $36.1$ to $82.3$ per cent (Table~S.VIII).

The jackknife removes the first-order bias, leaving an average absolute bias of $0.16$ and $0.11$ standard deviations of $\widehat\beta$ at the two sizes, but it is $1.37$ to $1.71$ and $1.30$ to $1.54$ times as dispersed as $\widehat\beta$. Because $\widehat\Omega$ estimates only the first-order variance (Theorem~\ref{thm:spj}), the intervals around $\widehat\beta_{\mathrm{SPJ}}$ with $\widehat\Omega$ cover only $66.9$ to $81.2$ and $76.2$ to $85.2$ per cent. The variance estimator $\widehat V_J$ of Theorem~\ref{thm:spjvariance} tracks the extra dispersion: in each of the $28$ configurations, the ratio of its mean standard error to that from $\widehat\Omega$ is $0.94$ to $1.01$ times the ratio of the Monte Carlo standard deviations of $\widehat\beta_{\mathrm{SPJ}}$ and $\widehat\beta$ (Table~S.XI), and its intervals cover $87.3$ to $92.8$ and $91.2$ to $95.6$ per cent. Its advantage comes from the dual values of the half-panel fits, which differ from those of the full panel at $3$ to $10$ per cent of the cells (Table~S.XII).

\newpage

\appendix

\noindent{\Large \textbf{Appendix}}

\section{Proof of Theorem~\ref{thm:slope}}\label{sec:proofslope}

This appendix proves Theorem~\ref{thm:slope}. The argument has four steps.

First, Appendix~\ref{ssec:additive} holds the slope fixed, or perturbs it by at most $O((NT)^{-1/2})$, and studies
the additive fit of the individual and time effects to the error array.
Lemma~\ref{lem:additive} bounds the fitted effects uniformly at the rate $\min(N,T)^{-1/3}$.
Lemma~\ref{lem:quadraticapprox} shows that they are close to an explicit weighted projection of
the quantile scores: each individual's average score divided by its density at zero, and each period's average
centred score across individuals divided by the average density. It also gives the improvement in the check loss from
fitting the effects. The later appendices use these lemmas as well.

Second, Appendix~\ref{ssec:rowscaling} writes the additive fit through linear-programming duality and proves the
weighted score identity of Lemma~\ref{lem:scaling}. Rescaling the errors of each individual by a factor that depends
on a weight makes the loss before the effects are fitted linear in the scaling, and the improvement in fit concave in
it, so the weighted sum that the slope expansion needs is a supergradient of a function that Lemma~\ref{lem:quadraticapprox}
evaluates. The two bias coefficients arise here: the individual part involves each individual's own density and the
time part the pooled density.

Third, Appendices~\ref{sssec:markedscore} and~\ref{sssec:localslope} decompose the subgradient of the profiled
criterion into a leading score, a term that is centred given the errors and negligible, and a term that
Lemma~\ref{lem:scaling} turns into the bias, and extend this decomposition to slopes within $O((NT)^{-1/2})$ of
$\beta_0$ by shifting each individual's errors by the movement of its quantile.

Fourth, Appendix~\ref{sssec:convexity} uses convexity. The local expansion of the subgradient gives a quadratic
approximation of the profiled criterion, uniformly on compact sets, which places every minimiser within
$\Op((NT)^{-1/2})$ of $\beta_0$ and yields \eqref{eq:slope}; the central limit theorem for independent triangular
arrays then gives \eqref{eq:slopeclt}.

\subsection{Weighted additive approximation under heterogeneous laws}\label{ssec:additive}

This subsection proves the growing-dimensional estimates used in all the
theorems. Let $W_{it}$ be independent, iid in $t$ within each individual, with
quantile-zero distribution function (CDF) $F_i^W$ and density $f_i^W$. Allow triangular laws, subject
to the common local bounds
\begin{equation}
F_i^W(0)=\tau,\qquad
0<c_*\le f_i^W(u)\le C_*,\qquad |(f_i^W)'(u)|\le C_*
\label{eq:scalarconditions}
\end{equation}
on a fixed neighbourhood of zero. Put $a_i^W=f_i^W(0)$,
$A^W=\sum_{i=1}^{N} a_i^W$, and $\bar a^W=A^W/N$.

Let $\A$ be the space of additive arrays $v_{it}=r_i+s_t$, with the
unique normalisation $\sum_{t=1}^{T}s_t=0$. Put
$\Lambda_W=\operatorname{diag}_{it}(a_i^W)\in\R^{NT\times NT}$ and
$\psi_W=(\psi(W_{it}))_{it}\in\R^{NT}$, where each weight $a_i^W$
appears $T$ times on the diagonal of $\Lambda_W$. Define the loss difference
\[
D_W(v):=\sum_{i,t}\{\rho(W_{it}-v_{it})-\rho(W_{it})\},\qquad v\in\A.
\]
For an arbitrary array of constants $u=(u_{it})_{it}$, $D_{W+u}$ is obtained
by replacing $W$ with $W+u$ in both loss terms. Define the minimiser set
\[
\M_W(u):=\arg\min_{v\in\A}D_{W+u}(v),
\]
which is nonempty because $\rho(e)\ge\min(\tau,1-\tau)|e|$ makes the continuous function $D_{W+u}$ coercive on $\A$.
For the baseline array $W$, define the quadratic approximation and its
unique minimiser by
\[
\widetilde D_W(v):=-v'\psi_W+\frac12v'\Lambda_Wv,\qquad
\widetilde v_W:=\arg\min_{v\in\A}\widetilde D_W(v).
\]
The corresponding optimised improvements are
\[
I_W:=-\min_{v\in\A}D_W(v),\qquad
\widetilde I_W:=-\min_{v\in\A}\widetilde D_W(v).
\]
The main results of this subsection are
Lemmas~\ref{lem:additive} and~\ref{lem:quadraticapprox}.
For every $v\in\M_W(u)$, with normalised decomposition
$v_{it}=r_i+s_t$, the first bounds
$\norm{v}_2$ and $\max_i|r_i|+\max_t|s_t|$, uniformly over
small perturbations. The second bounds
$\norm{v-\widetilde v_W}_2^2$ uniformly over the same
perturbations. At $u=0$, it also bounds $|I_W-\widetilde I_W|$ and
gives an explicit deterministic leading term for $\widetilde I_W$,
and hence for $I_W$.

Fix $0\le L<\infty$ and put $d_{NT}=L/\sqrt{NT}$.
All probabilistic bounds in this subsection hold uniformly over triangular
individual laws satisfying \eqref{eq:scalarconditions} with common constants.
The bounds for the perturbed criterion $D_{W+u}$ and its minimisers hold
uniformly over $\norm u_\infty\le d_{NT}$; no condition is placed on how the
$u_{it}$ vary across $(i,t)$. They therefore also apply
to a random array, possibly dependent on the data, that, for every $\epsilon>0$ and some $L=L_\epsilon$, lies in this ball with probability at least $1-\epsilon$ for large $N,T$, such as
$u_{it}=-x_{it}'(\widehat\beta-\beta_0)$ in the proof of Lemma~\ref{lem:uniformrate}.
Lemmas~\ref{lem:additive} and~\ref{lem:quadraticapprox} use the growth
condition in A4 only through $\log(NT)=o(\min(N,T)^{1/3})$.

\subsubsection{Geometry}\label{sssec:geometry}
For any array $w$, write
\[
\bar w_{i\cdot}=T^{-1}\sum_t w_{it},\qquad
\bar w_{\cdot t}=N^{-1}\sum_i w_{it},\qquad
\bar w_{\cdot\cdot}=(NT)^{-1}\sum_{i,t}w_{it}.
\]
The additive space $\A$ has dimension $N+T-1$. Its Euclidean projection
$P$ has diagonal entries $(N+T-1)/(NT)$, and
\begin{equation}
(Pw)_{it}=\bar w_{i\cdot}+\bar w_{\cdot t}-\bar w_{\cdot\cdot},
\qquad \norm v_2^2=T\sum_{i=1}^{N}r_i^2+N\sum_{t=1}^{T}s_t^2
\quad(v\in\A).
\label{eq:geometry}
\end{equation}
Applying Lemma~\ref{lem:weightedquadratic} in
Appendix~\ref{ssec:quadraticgeometry} with $a_i=a_i^W$ and $g=\psi_W$
gives the quadratic minimiser
\[
\widetilde v_W=M_W\psi_W,\qquad
M_W=B(B'\Lambda_WB)^{-1}B',
\]
where $B$ is any basis matrix of $\A$. The matrix $M_W$ is deterministic
and symmetric; $M_W\Lambda_W$ is the weighted projection onto $\A$.
Writing $\bar\psi_{i\cdot}^W=T^{-1}\sum_t\psi(W_{it})$, we have
\begin{equation}
(\widetilde v_W)_{it}=\frac{\bar\psi_{i\cdot}^W}{a_i^W}
+\frac1{A^W}\sum_{j=1}^{N}\{\psi(W_{jt})-\bar\psi_{j\cdot}^W\}.
\label{eq:weightedprojection}
\end{equation}
The same lemma gives $M_W\Lambda_WM_W=M_W$ and, for $v\in\A$,
\begin{equation}
\widetilde D_W(v)
=\frac12(v-\widetilde v_W)'\Lambda_W(v-\widetilde v_W)
-\frac12\psi_W'M_W\psi_W.
\label{eq:completequadratic}
\end{equation}
Thus excess quadratic loss is a weighted squared distance from the
minimiser, and $\widetilde I_W=\psi_W'M_W\psi_W/2$.

The individual-constant and common time-contrast subspaces are
orthogonal; the indicator of individual $i$ is an eigenvector of $M_W$ with eigenvalue $1/a_i^W$,
every common time contrast is one with eigenvalue $1/\bar a^W$, and
$M_W$ vanishes on the orthogonal complement of the two subspaces.
Lemma~\ref{lem:weightedtraces} gives the corresponding trace identities,
\begin{equation}
\begin{split}
\operatorname{tr}M_W
&=\sum_{i=1}^{N}\frac1{a_i^W}+\frac{T-1}{\bar a^W},\\
\operatorname{tr}(M_W^2)
&=\sum_{i=1}^{N}\frac1{(a_i^W)^2}+\frac{T-1}{(\bar a^W)^2}.
\end{split}
\label{eq:traces}
\end{equation}
The two terms correspond to the $N$ individual directions and the
$T-1$ common time-contrast directions. Since $a_i^W\ge c_*$,
$\operatorname{tr}(M_W^k)\le c_*^{-k}(N+T-1)=O(N+T)$ for $k=1,2$.

\subsubsection{Empirical-process and empirical-CDF bounds}\label{sssec:bounds}
Set
\[
R(w,t)=\int_0^t\{\ind(w\le s)-\ind(w\le0)\}\,ds.
\]
Symmetrisation (\citet[Lemma~2.3.6]{vdvw}) and scalar Rademacher contraction
(\citet[Proposition~A.3.2]{vdvw}; \citet[Theorem~4.12]{ledouxtalagrand}) give, for every $r>0$,
\begin{equation}
\E\sup_{v\in\A,\ \norm v_2\le r}
|D_W(v)-\E D_W(v)|\le Cr\sqrt{N+T}.
\label{eq:globalprocess}
\end{equation}
Here is the projection calculation underlying this and the needed local
version. In vector expressions, indices $j,\ell$ range over the $NT$ cells
in a fixed order. For independent Rademacher signs $\sigma=(\sigma_j)_j$, independent of $W$,
and coordinatewise Lipschitz constants $\mathbf L=(L_j)_j$, contraction
bounds the centred supremum by
\[
Cr\E\norm{P(\sigma\odot\mathbf L)}_2
\le Cr\left(\sum_jP_{jj}\E L_j^2\right)^{1/2}.
\]
For loss differences, $L_j=\max(\tau,1-\tau)$. For $R(W_j,t)$ on
$|t|\le\delta$, with $\delta>0$ sufficiently small,
one can use $L_j=\ind\{|W_j|\le\delta\}$, whose expectation is
uniformly at most $C\delta$. Consequently,
\begin{equation}
\E\sup_{\substack{v\in\A,\ \norm v_2\le r\\\norm v_\infty\le\delta}}
\left|\sum_j\{R(W_j,v_j)-\E R(W_j,v_j)\}\right|
\le Cr\sqrt{\delta{(N+T)}}.
\label{eq:localprocess}
\end{equation}
The scalar derivative bound in \eqref{eq:scalarconditions} also gives
\begin{equation}
\left|\sum_{i,t}\E R(W_{it},v_{it})
-\frac12\sum_{i,t}a_i^W v_{it}^2\right|
\le C\delta\norm v_2^2\quad(\norm v_\infty\le\delta).
\label{eq:meanremainder}
\end{equation}
This follows from
$\E R(W_{it},s)=\int_0^s\{F_i^W(u)-\tau\}\,du
=a_i^Ws^2/2+O(|s|^3)$, uniformly in $i$.
Only bounded loss differences are integrated, so heavy tails cause no
integrability issue.

Let $F_{i\cdot,T}$ be the empirical CDF in individual $i$, let
$F_{\cdot t,N}$ be the empirical CDF in time $t$, and put
$\bar F_N^W=N^{-1}\sum_{i=1}^{N}F_i^W$. Then
\begin{equation}
\Delta_{NT}:=\max\left\{
\max_i\sup_u|F_{i\cdot,T}(u)-F_i^W(u)|,
\max_t\sup_u|F_{\cdot t,N}(u)-\bar F_N^W(u)|\right\}
=\Op\!\left(\sqrt{\frac{\log(NT)}{\min(N,T)}}\right).
\label{eq:cdf}
\end{equation}
For individuals, the Dvoretzky--Kiefer--Wolfowitz inequality (\citet{dkw}), with the
constant of \citet{massart}, and a union bound
give $2N\exp(-2Ts^2)$ as an upper bound on exceedance of $s$.
For columns, iid sampling across $i$ is not available. Instead take a
grid of $(NT)^2$ quantiles of the continuous mixture CDF $\bar F_N^W$.
At each grid point, Hoeffding's inequality (\citet{hoeffding}) applies to the independent,
possibly nonidentical indicators in a column. A union bound over columns
and grid points gives $2T(NT)^2\exp(-2Ns^2)$. Monotonicity bounds the
off-grid discrepancy by the maximum grid discrepancy plus $(NT)^{-2}$.
This proves \eqref{eq:cdf}, including for left empirical-CDF limits.

\subsubsection{Uniform localisation of additive minimisers}\label{sssec:localization}

\begin{lemma}[Uniform localisation of additive minimisers]\label{lem:additive}
Under \eqref{eq:scalarconditions} and the growth condition of A4,
uniformly over $\norm u_\infty\le d_{NT}$ and all minimisers
$v\in\M_W(u)$, with the suprema over $u$ and $v$ inside the probability,
\begin{align}
\norm{v}_2&=\Op(\sqrt{N+T}),\label{eq:nuisanceradius}\\
\max_i|r_i|+\max_t|s_t|&=\Op(\min(N,T)^{-1/3}),
\label{eq:nuisancemax}
\end{align}
where $v_{it}=r_i+s_t$ is the normalised additive decomposition.
The growth condition of A4 enters in Step~2 of the proof, where it gives
$\Delta_{NT}=o_p(\min(N,T)^{-1/3})$.
\end{lemma}
\begin{proof}
The proof has four steps. In Steps~1--3, all bounds are simultaneous
over the permitted perturbations and all constrained minimisers.

\textbf{1. Constrained minimiser norm bound.}
Choose a small fixed $\delta_0>0$ and initially minimise over the compact
convex set
\[
\A_{\delta_0}
=\{v\in\A:\max_i|r_i(v)|\le\delta_0,\ \max_t|s_t(v)|\le\delta_0\},
\]
with $\delta_0$ small enough that all thresholds below lie in the common
density neighbourhood. Fix $\norm u_\infty\le d_{NT}$ and choose any
constrained minimiser
\[
v^\circ\in\arg\min_{v\in\A_{\delta_0}}D_{W+u}(v).
\]
Write $v^\circ_{it}=r_i^\circ+s_t^\circ$, with $\sum_t s_t^\circ=0$.
For any $v\in\A_{\delta_0}$, decompose
\[
D_{W+u}(v)
=\E D_W(v)+\{D_W(v)-\E D_W(v)\}
+\{D_{W+u}(v)-D_W(v)\}.
\]
Knight's identity and $F_i^W(0)=\tau$ give
\[
\E D_W(v)
=\sum_{i,t}v_{it}^2\int_0^1(1-s)f_i^W(sv_{it})\,ds
\ge\frac{c_*}{2}\norm v_2^2.
\]
Also, writing $\rho_\tau(x)=|x|/2+(\tau-1/2)x$, the linear terms cancel
in the perturbation difference, and the reverse triangle inequality gives
\[
|D_{W+u}(v)-D_W(v)|
\le\sum_{i,t}|u_{it}|\le NT\,d_{NT}=O(N+T).
\]
By optimality, $D_{W+u}(v^\circ)\le0$. If $\norm{v^\circ}_2\ge r>0$,
define
\[
v_r=\frac{r}{\norm{v^\circ}_2}v^\circ\in\A_{\delta_0},
\qquad \norm{v_r}_2=r.
\]
Since $D_{W+u}(0)=0$, convexity gives
\[
D_{W+u}(v_r)
\le\frac{r}{\norm{v^\circ}_2}D_{W+u}(v^\circ)\le0.
\]
Hence
\[
\sup_{v\in\A,\ \norm v_2\le r}|D_W(v)-\E D_W(v)|
\ge(c_*/2)r^2-NT\,d_{NT}.
\]
For a positive right-hand side, \eqref{eq:globalprocess} and Markov's
inequality bound the probability that any permitted constrained
minimiser has norm at least $r$ by
\[
\frac{Cr\sqrt{N+T}}{(c_*/2)r^2-NT\,d_{NT}}.
\]
Taking $r=M\sqrt{N+T}$ and then $M$ large proves the uniform
$\Op(\sqrt{N+T})$ bound. In particular, \eqref{eq:geometry} gives
\begin{equation}
\frac1T\sum_{t=1}^{T}(s_t^\circ)^2
\le\frac{\norm{v^\circ}_2^2}{NT}
=\Op(N^{-1}+T^{-1})=\Op(\min(N,T)^{-1}).
\label{eq:averagesquares}
\end{equation}

\textbf{2. Individual bounds from one-sided quantile inequalities.}
For $\eta>0$, let
\[
b_\eta:=T^{-1}\sum_t\ind\{|s_t^\circ|\ge\eta\}
\le\frac{T^{-1}\sum_t (s_t^\circ)^2}{\eta^2}.
\]
If $r_i^\circ>\eta+d_{NT}$, then $W_{it}<r_i^\circ-\eta-d_{NT}$ implies a strictly
negative residual whenever $|s_t^\circ|<\eta$. Thus
\begin{align*}
\frac1T\sum_t\ind\{W_{it}+u_{it}<r_i^\circ+s_t^\circ\}
&\ge\frac1T\sum_t\ind\{W_{it}<r_i^\circ-\eta-d_{NT}\}-b_\eta\\*
&\ge F_i^W(r_i^\circ-\eta-d_{NT})-\Delta_{NT}-b_\eta.
\end{align*}
By optimality of $v^\circ$, the fraction on the left cannot exceed $\tau$:
otherwise decreasing the positive $r_i^\circ$ slightly would be feasible and
reduce the loss. Since $F_i^W(0)=\tau$, the lower density bound gives
\[
r_i^\circ\le\eta+d_{NT}+c_*^{-1}(\Delta_{NT}+b_\eta).
\]
The positive-residual inequality, with threshold $1-\tau$, gives the
same bound for $-r_i^\circ$. Therefore
\begin{equation}
\max_i|r_i^\circ|
\le\eta+d_{NT}+C\left\{\Delta_{NT}
+\frac{T^{-1}\sum_t (s_t^\circ)^2}{\eta^2}\right\}.
\label{eq:rowmaxbound}
\end{equation}
Take $\eta=\min(N,T)^{-1/3}$ and use \eqref{eq:cdf} and
\eqref{eq:averagesquares}. Under A4,
$\Delta_{NT}=o_p(\min(N,T)^{-1/3})$, so
$R_*:=\max_i|r_i^\circ|=\Op(\min(N,T)^{-1/3})$.

\textbf{3. Time bounds with the centring constraint.}
If $s_t^\circ>0$, choose $t'$ with $s_{t'}^\circ\le0$, which is possible because $\sum_t s_t^\circ=0$.
Holding the individual effects $r_i^\circ$ fixed, write the time-$t$ loss as
\[
d_t(s):=\sum_i\{\rho(W_{it}+u_{it}-r_i^\circ-s)-\rho(W_{it}+u_{it})\}.
\]
Its left derivative at $s_t^\circ$ satisfies
\begin{align*}
d'_{t,-}(s_t^\circ)
&=\sum_i\ind\{W_{it}+u_{it}<r_i^\circ+s_t^\circ\}-N\tau\\
&\ge\sum_i\ind\{W_{it}<s_t^\circ-R_*-d_{NT}\}-N\tau\\
&\ge N\{\bar F_N^W(s_t^\circ-R_*-d_{NT})-\tau-\Delta_{NT}\}.
\end{align*}
The right derivative for time $t'$, using $s_{t'}^\circ\le0$, satisfies
\begin{align*}
d'_{t',+}(s_{t'}^\circ)
&=\sum_i\ind\{W_{it'}+u_{it'}\le r_i^\circ+s_{t'}^\circ\}-N\tau\\
&\le\sum_i\ind\{W_{it'}\le R_*+d_{NT}\}-N\tau\\
&\le N\{\bar F_N^W(R_*+d_{NT})-\tau+\Delta_{NT}\}.
\end{align*}
When $s_t^\circ>2(R_*+d_{NT})$, the lower density bound yields
\[
d'_{t,-}(s_t^\circ)-d'_{t',+}(s_{t'}^\circ)
\ge N\{c_*[s_t^\circ-2(R_*+d_{NT})]-2\Delta_{NT}\}.
\]
If this were positive, decreasing $s_t^\circ$ and increasing $s_{t'}^\circ$ equally
would preserve the centring constraint and strictly reduce the loss.
Thus $s_t^\circ\le2(R_*+d_{NT})+2\Delta_{NT}/c_*$. For $s_t^\circ<0$, choose
$s_{t'}^\circ\ge0$ and reverse the feasible change to obtain the same bound
for $-s_t^\circ$. Consequently,
\[
\max_t|s_t^\circ|
\le2(R_*+d_{NT})+2\Delta_{NT}/c_*
=\Op(\min(N,T)^{-1/3}).
\]
The one-sided derivatives include zero residuals, and the paired
changes remain feasible at the boundary of $\A_{\delta_0}$.

\textbf{4. Removal of the constraint.}
The preceding bounds place every constrained minimiser $v^\circ$ in the relative
interior of $\A_{\delta_0}$, simultaneously for all permitted $u$, with
probability tending to one. Each is then an unconstrained local, hence
global, optimum on $\A$. An unconstrained optimum outside
$\A_{\delta_0}$ would, by convexity, produce a boundary optimum along
its segment to an interior optimum, contradicting the bounds.
Thus every unconstrained minimiser $v\in\M_W(u)$ satisfies
\eqref{eq:nuisanceradius}--\eqref{eq:nuisancemax}.
\end{proof}

\subsubsection{Quadratic approximation of improvement}\label{sssec:quadraticapprox}

\begin{lemma}[Quadratic approximation of improvement]\label{lem:quadraticapprox}
Under the assumptions and notation of Lemma~\ref{lem:additive},
uniformly over $\norm u_\infty\le d_{NT}$ and all minimisers
$v\in\M_W(u)$, with the suprema over $u$ and $v$ inside the probability,
\begin{equation}
\norm{v-\widetilde v_W}_2^2
=\Op\{{(N+T)}\min(N,T)^{-1/6}\}.
\label{eq:nuisanceapprox}
\end{equation}
For the unperturbed improvement $I_W$,
\begin{equation}
\begin{split}
I_W&=\widetilde I_W+\Op\{{(N+T)}\min(N,T)^{-1/6}\}\\
&=\frac{\tau(1-\tau)}{2}\left\{\sum_{i=1}^{N}\frac1{a_i^W}+\frac{T-1}{\bar a^W}\right\}
+\Op\{{(N+T)}\min(N,T)^{-1/6}\}.
\end{split}
\label{eq:improvement}
\end{equation}
\end{lemma}
\begin{proof}
Fix positive constants $M,C$ and put
$r=M\sqrt{N+T}$ and $\delta=C\min(N,T)^{-1/3}$.
Knight's identity,
$\rho(w-v)-\rho(w)=-v\psi(w)+R(w,v)$, and
\eqref{eq:localprocess}--\eqref{eq:meanremainder} give
\begin{equation}
\begin{split}
D_W(v)-\widetilde D_W(v)
&=\sum_{i,t}R(W_{it},v_{it})-\frac12\sum_{i,t}a_i^Wv_{it}^2\\
&=\Op\{(N+T)\min(N,T)^{-1/6}\},
\end{split}
\label{eq:quadratic}
\end{equation}
uniformly on $v\in\A$, $\norm v_2\le r$, $\norm v_\infty\le\delta$.
The expected centred supremum is $O(r\sqrt{\delta(N+T)})$, whereas
the deterministic remainder is $O(\delta r^2)$, of the smaller order
$O\{(N+T)\min(N,T)^{-1/3}\}$.

On the same set, each cell's perturbation difference is bounded by
$|u_{it}|$ and vanishes when $|W_{it}|>|v_{it}|+|u_{it}|$. Hence
\begin{equation}
\begin{split}
\sup_{\norm u_\infty\le d_{NT}}|D_{W+u}(v)-D_W(v)|
&\le d_{NT}\sum_{i,t}\ind\{|W_{it}|\le\delta+d_{NT}\}\\
&=\Op\{(N+T)\min(N,T)^{-1/3}\}.
\end{split}
\label{eq:fineperturb}
\end{equation}
The last bound follows from Markov's inequality, since the expectation
is $O(NT\,d_{NT}\delta)=O\{(N+T)\min(N,T)^{-1/3}\}$.

The entries $\psi_j=(\psi_W)_j$ are independent centred variables taking
values $\tau$ and $\tau-1$ with probabilities $1-\tau$ and $\tau$.
Consequently
\[
\E\psi_j^2=\tau(1-\tau),\qquad
\E\psi_j^4=\tau(1-\tau)\{1-3\tau(1-\tau)\}.
\]
Since $\widetilde v_W=M_W\psi_W$ and $M_W$ is deterministic and symmetric,
\[
\E\norm{\widetilde v_W}_2^2
=\E(\psi_W'M_W^2\psi_W)
=\tau(1-\tau)\operatorname{tr}(M_W^2)
=O(N+T),
\]
where the last bound follows from \eqref{eq:traces}.
Markov's inequality now gives
$\norm{\widetilde v_W}_2=\Op(\sqrt{N+T})$.
By Hoeffding's inequality and \eqref{eq:weightedprojection},
$\norm{\widetilde v_W}_\infty
=\Op(\sqrt{\log(NT)/\min(N,T)})$.
Since $\log(NT)=o(\min(N,T)^{1/3})$, these bounds together with
Lemma~\ref{lem:additive} place both $\widetilde v_W$ and every
$v\in\M_W(u)$ in the displayed neighbourhoods with
arbitrarily high probability by increasing the fixed constants $M,C$.

For any such minimiser $v$, optimality gives
\[
D_{W+u}(v)\le D_{W+u}(\widetilde v_W).
\]
Combining this inequality with \eqref{eq:quadratic},
\eqref{eq:fineperturb}, and \eqref{eq:completequadratic} yields
\[
\frac{c_*}{2}\norm{v-\widetilde v_W}_2^2
\le \widetilde D_W(v)-\widetilde D_W(\widetilde v_W)
=\Op\{{(N+T)}\min(N,T)^{-1/6}\}.
\]
This proves \eqref{eq:nuisanceapprox} with the stated uniformity.
For $u=0$, on the event that both minimisers belong to the displayed
neighbourhood, their minimum values differ by at most the uniform
criterion error:
\[
|I_W-\widetilde I_W|
\le
\sup_{\substack{v\in\A,\ \norm v_2\le r\\\norm v_\infty\le\delta}}
|D_W(v)-\widetilde D_W(v)|.
\]
By \eqref{eq:quadratic} and the preceding localisation, this is
$\Op\{(N+T)\min(N,T)^{-1/6}\}$, proving the first equality in
\eqref{eq:improvement}.

Independence also gives
\begin{align*}
\E(\psi_W'M_W\psi_W)
&=\operatorname{tr}\{M_W\E(\psi_W\psi_W')\}
=\tau(1-\tau)\operatorname{tr}(M_W),\\
\operatorname{Var}(\psi_W'M_W\psi_W)
&=2(\tau(1-\tau))^2\sum_{j\ne\ell}(M_W)_{j\ell}^2
 +\tau(1-\tau)\{1-4\tau(1-\tau)\}\sum_j(M_W)_{jj}^2=O{(N+T)}.
\end{align*}
To verify the variance formula, separate the diagonal and off-diagonal
parts. Their covariance is zero because a remaining single score has
mean zero. Distinct unordered off-diagonal pairs are uncorrelated, and
$\operatorname{Var}(\psi_j^2)=\tau(1-\tau)\{1-4\tau(1-\tau)\}$.
The diagonal fluctuation vanishes only at $\tau=1/2$. 

By \eqref{eq:traces} and Chebyshev's inequality,
\[
\widetilde I_W
=\frac{\tau(1-\tau)}{2}\left\{\sum_{i=1}^{N}\frac1{a_i^W}
+\frac{T-1}{\bar a^W}\right\}+\Op(\sqrt{N+T}).
\]
Since $\sqrt{N+T}=O\{(N+T)\min(N,T)^{-1/6}\}$, this proves the
second equality in \eqref{eq:improvement}.
All bounds use common constants, so the stated uniformity follows.
Suprema can be taken over measurable finite-dimensional sets: the
criteria are continuous and minimisers for bounded $u$ lie in a finite
data-dependent ball by the triangle inequality.
\end{proof}

\subsection{An individual-scaling identity for the heterogeneous bias}\label{ssec:rowscaling}

Let
\[
\D:=\{q\in\R^{N\times T}:Pq=0,\ \tau-1\le q_{it}\le\tau\text{ for all }i,t\}.
\]
The condition $Pq=0$ is equivalent to zero individual and time sums.
Applying Lemma~\ref{lem:quantileduality} in Appendix~\ref{ssec:quantileduality}
to the additive space $\A$ gives, with both optima attained,
\begin{equation}
\min_{v\in\A}\sum_{i,t}\rho(w_{it}-v_{it})=\max_{q\in\D}w'q.
\label{eq:dual}
\end{equation}
Take the measurable maximising selection $q(w)$ furnished by the lemma.
It depends only on $w$ and is unchanged by additive shifts.
For any primal minimiser $\widehat v$, complementary slackness and orthogonality give
\[
q_{it}(w)\in\partial\rho(w_{it}-\widehat v_{it}),\qquad \widehat v'q(w)=0.
\]
For a fixed array $w$, use the improvement $I_w$ defined in
Appendix~\ref{ssec:additive}. Since $\rho(w_{it})=w_{it}\psi(w_{it})$, the exact identities
\begin{equation}
\sum_{i,t}w_{it}\{q_{it}(w)-\psi(w_{it})\}=-I_w,\qquad
\sum_{i,t}|w_{it}|\,|q_{it}(w)-\psi(w_{it})|=I_w
\label{eq:exact}
\end{equation}
follow. The second equality holds because every summand in the first is
nonpositive. Also,
\begin{equation}
q_{it}(w)\ne\psi(w_{it})\quad\Longrightarrow\quad
|w_{it}|\le\norm{\widehat v}_\infty.
\label{eq:support}
\end{equation}
Fitting the effects can therefore change a score only near zero.
Indeed, if $|w_{it}|>\norm{\widehat v}_\infty$, the original observation and fitted residual
have the same nonzero sign, so complementary slackness gives $q_{it}(w)=\psi(w_{it})$.
For the random array $W$, Lemma~\ref{lem:additive} makes this neighbourhood shrink to zero, and
the bounded local densities imply that only a vanishing fraction of scores change.
Together with \eqref{eq:exact}, this localisation also controls the remainder from a local expansion
of the conditional regressor mean in Appendix~\ref{sssec:markedscore}.

\begin{lemma}[Weighted score identity]\label{lem:scaling}
Under \eqref{eq:scalarconditions} and the growth condition of A4, let
$\zeta=(\zeta_1,\ldots,\zeta_N)'$ be any deterministic vector of individual weights, bounded
uniformly in $i$, $N$ and $T$. Then
\begin{equation}
\begin{split}
J_W(\zeta)&:=-\sum_{i,t}\zeta_iW_{it}
\{q_{it}(W)-\psi(W_{it})\}\\
&=\frac{\tau(1-\tau)}{2}\left\{\sum_{i=1}^{N}\frac{\zeta_i}{a_i^W}
+(T-1)\frac{N^{-1}\sum_{i=1}^{N} a_i^W\zeta_i}{(\bar a^W)^2}\right\}
+\Op\{{(N+T)}\min(N,T)^{-1/12}\}.
\end{split}
\label{eq:weightedidentity}
\end{equation}
The $\Op$ constant depends only on a bound on the weights and the
constants in \eqref{eq:scalarconditions}, so the assertion is uniform over
bounded deterministic weight sequences and triangular scalar laws with
common constants in this sense. It is not a bound on the supremum over
weights inside the probability, which would be of larger order because
the error is linear in $\zeta$; only finitely many weight sequences are used below.
\end{lemma}

\begin{proof}
Choose a fixed $s_0>0$ with $s_0\sup_i|\zeta_i|\le1/2$ uniformly in $N,T$,
and set $W_{it}(s)=(1+s\zeta_i)W_{it}$ for $|s|\le s_0$.
Positive scaling preserves independence, the within-individual iid property,
and zero as the $\tau$-quantile. The density is
\[
f_{i,s}(u)=\frac{1}{1+s\zeta_i}
f_i^W\!\left(\frac{u}{1+s\zeta_i}\right).
\]
Thus \eqref{eq:scalarconditions} holds on half the original neighbourhood,
uniformly in $i,s,N,T$, with constants $2c_*/3$ and $4C_*$.
The density at zero is $a_i^W/(1+s\zeta_i)$.

Using the improvement $I_W$ and criterion $D_W$ defined in
Appendix~\ref{ssec:additive}, set
\[
I_W(s):=I_{W(s)}=-\min_{v\in\A}D_{W(s)}(v).
\]
By \eqref{eq:dual} and positive homogeneity of $\rho$,
\[
I_W(s)=\min_{q\in\D}\sum_{i,t}(1+s\zeta_i)
\{\rho(W_{it})-W_{it}q_{it}\}.
\]
Since $\D$ does not depend on $s$, this is a pointwise minimum of affine
functions of $s$, and hence is concave
\citep[Section~3.2.3, p.~81]{boydvandenberghe2004}.
Moreover, $q(W)$ attains this minimum at $s=0$.
Evaluating the objective at this same $q(W)$ for other values of $s$ gives
\[
\begin{aligned}
I_W(s)
&\le \sum_{i,t}(1+s\zeta_i)
\{\rho(W_{it})-W_{it}q_{it}(W)\}
=I_W(0)+sJ_W(\zeta),
\end{aligned}
\]
where the equality uses $\rho(W_{it})=W_{it}\psi(W_{it})$.
Thus $J_W(\zeta)$ is a supergradient of $I_W(s)$ at zero.
Taking $s=h$ and $s=-h$, for $0<h\le s_0$, yields
\begin{equation}
\frac{I_W(h)-I_W(0)}h\le J_W(\zeta)
\le\frac{I_W(0)-I_W(-h)}h.
\label{eq:secants}
\end{equation}
Applying Lemma~\ref{lem:quadraticapprox} to these scaled laws gives,
also for deterministic sequences $s\in[-s_0,s_0]$,
\begin{equation}
\begin{split}
I_W(s)&=I^*_{NT}(s)+\Op\{{(N+T)}\min(N,T)^{-1/6}\},\\
I^*_{NT}(s) &:= \frac{\tau(1-\tau)}{2}\left\{
\sum_{i=1}^{N}\frac{1+s\zeta_i}{a_i^W}
+\frac{T-1}{N^{-1}\sum_{i=1}^{N} a_i^W/(1+s\zeta_i)}\right\}.
\end{split}
\label{eq:scaledimprovement}
\end{equation}
Direct differentiation gives the leading expression in
\eqref{eq:weightedidentity} as $(I^*_{NT})'(0)$, and
$\sup_{|s|\le s_0}|(I^*_{NT})''(s)|\le C{(N+T)}$.
Insert \eqref{eq:scaledimprovement} into \eqref{eq:secants} with
$h=\min(N,T)^{-1/12}$, which is at most $s_0$ for large $N,T$. The deterministic secant error and the stochastic error are
\[
O\{{(N+T)}h\}\qquad\text{and}\qquad\Op\{{(N+T)}\min(N,T)^{-1/6}/h\},
\]
respectively. Both are $O_p\{{(N+T)}\min(N,T)^{-1/12}\}$, proving the claim. This proof uses sample loss
differences only; no expectation of the raw absolute loss is needed.
\end{proof}

\subsection{Profile score and proof of the slope theorem}\label{ssec:profile}

\subsubsection{A marked-array score expansion}\label{sssec:markedscore}
Consider independent pairs $(x_{it},W_{it})$, iid over $t$ within each
individual, with bounded $x$ and scalar laws satisfying
\eqref{eq:scalarconditions}. Suppose near zero the conditional mark mean
has the uniform expansion
\begin{equation}
m_i^W(u):=\E_i[x\mid W=u]
=\mu_i^W+\nu_i^Wu+r_i^W(u),\qquad
\norm{r_i^W(u)}\le |u|\omega(|u|),\qquad
\sup_i\norm{\nu_i^W}<\infty,
\label{eq:markexpansion}
\end{equation}
where $\omega(\delta)\to0$. This condition is verified from A2 below.
Take any $\widehat v\in\M_W(0)$.
Zero individual sums of $q(W)$ imply
\begin{align}
S_W:=\sum_{i,t}x_{it}q_{it}(W)
={}&\sum_{i,t}(x_{it}-\mu_i^W)\psi(W_{it})\notag\\
&+\sum_{i,t}\{x_{it}-m_i^W(W_{it})\}
\{q_{it}(W)-\psi(W_{it})\}\notag\\
&+\sum_{i,t}\{m_i^W(W_{it})-\mu_i^W\}
\{q_{it}(W)-\psi(W_{it})\}.
\label{eq:scoredecomp}
\end{align}
Conditional on the complete $W$ array, the terms on the second line
are independent and mean zero. Indeed, the cells are independent and
$q(W)$ is a function of $W$, so given $W$ the $x_{it}$ are independent
with $\E[x_{it}\mid W]=\E_i[x\mid W=W_{it}]=m_i^W(W_{it})$. Their conditional squared-norm variance
is bounded by
\[
4K^2\sum_{i,t}|q_{it}(W)-\psi(W_{it})|^2.
\]
By \eqref{eq:support} and \eqref{eq:nuisancemax}, this sum divided by
$NT$ converges to zero in probability. Indeed, after restricting to
$\norm{\widehat v}_\infty\le\delta$, the expected fraction of possible
changes is at most $C\delta$; then let $\delta\downarrow0$.
Conditional Chebyshev with its bound truncated at one proves that the
second line is $\op(\sqrt{NT})$.

For the third line, take $\omega$ nondecreasing (replacing it by its
running supremum if necessary). By \eqref{eq:support}, $q_{it}(W)\ne\psi(W_{it})$
only if $|W_{it}|\le\norm{\widehat v}_\infty$, so the remainder in
\eqref{eq:markexpansion} contributes at most
\[
\omega(\norm{\widehat v}_\infty)\sum_j|W_j|\,|q_j(W)-\psi(W_j)|
=\omega(\norm{\widehat v}_\infty)I_W
\]
by \eqref{eq:exact}, which is $\op{(N+T)}$
by \eqref{eq:nuisancemax} and \eqref{eq:improvement}. Apply Lemma~\ref{lem:scaling} coordinatewise to the
bounded vector weights $\nu_i^W$. Since $p$ is fixed, it follows that
\begin{equation}
\begin{split}
S_W={}&\sum_{i,t}(x_{it}-\mu_i^W)\psi(W_{it})\\
&-\frac{\tau(1-\tau)}{2}\left\{\sum_{i=1}^{N}\frac{\nu_i^W}{a_i^W}
+(T-1)\frac{N^{-1}\sum_{i=1}^{N} a_i^W\nu_i^W}{(\bar a^W)^2}\right\}
+\op(\sqrt{NT}).
\end{split}
\label{eq:markedscore}
\end{equation}
Here $N+T\asymp\sqrt{NT}$ and ${(N+T)}\min(N,T)^{-1/12}=o(\sqrt{NT})$.
This establishes the heterogeneous score expansion without any
identical-distribution assumption across individuals.

\subsubsection{Deterministic local slope displacements}\label{sssec:localslope}
After subtracting the true effects and writing $b=\beta-\beta_{0}$, define
\begin{equation}
Q_{NT}(b)=\min_{v\in\A}\sum_{i,t}\rho(\varepsilon_{it}-x_{it}'b-v_{it})
=\max_{q\in\D}(\varepsilon-Xb)'q.
\label{eq:profile}
\end{equation}
Here $(Xb)_{it}=x_{it}'b$ and $X'q=\sum_{i,t}x_{it}q_{it}$.
A valid subgradient is $S_{NT}(b)=-X'q(\varepsilon-Xb)$.
For a small deterministic $b$, let $\xi_i(b)$ be the individual-$i$ marginal
quantile of $\varepsilon-x'b$, and set
\[
W_{it,b}=\varepsilon_{it}-x_{it}'b-\xi_i(b).
\]
Write $W_b=\varepsilon-x'b-\xi_i(b)$ for a generic draw from individual $i$.
The conditional quantile restriction and the positive conditional densities at zero in A2 first give
$|\xi_i(b)|\le K\norm b$. Uniform Taylor expansion of the quantile
equation then gives
\begin{equation}
\xi_i(b)=-\mu_i'b+O(\norm b^2)
\label{eq:quantilemovement}
\end{equation}
uniformly over individuals. The marginal density of $W_{it,b}$ and its conditional mark mean are
\begin{align}
f_{i,b}(u)&=\E_i f_i(u+\xi_i(b)+x'b\mid x),\notag\\
m_{i,b}(u)&=
\frac{\E_i[xf_i(u+\xi_i(b)+x'b\mid x)]}{f_{i,b}(u)}.
\label{eq:bayes}
\end{align}
Their denominators are uniformly positive on a fixed smaller neighbourhood.
More precisely, put $r_1=\min\{r_0,c_0/(2C_0)\}$, so that $c_0/2\le f_i(u\mid x)\le C_0$ for $|u|\le r_1$
by A2, and $b_0=r_1/(4K+2)$. For
$\norm b\le b_0$ the individual laws of $W_{it,b}$ satisfy \eqref{eq:scalarconditions}
on $[-r_1/2,r_1/2]$ with $c_0/2$ and $C_0$ in place of $c_*$ and $C_*$, so that
Lemmas~\ref{lem:additive}, \ref{lem:quadraticapprox}, and~\ref{lem:scaling} and \eqref{eq:markedscore}
apply to them with common constants. By Bayes' formula, $m_{i,b}$ is a
version of $\E_i[x\mid W_b=u]$.
Writing $a_{i,b}=f_{i,b}(0)$, $\mu_{i,b}=m_{i,b}(0)$,
and $\nu_{i,b}=m_{i,b}'(0)$, A2 yields uniformly in $i$,
\begin{equation}
a_{i,b}=a_i+O(\norm b),\quad
\mu_{i,b}=\mu_i+O(\norm b),\quad
\nu_{i,b}=\frac{c_i}{a_i}+o(1).
\label{eq:shiftedcoefficients}
\end{equation}
For the derivative assertion, differentiate the quotient in
\eqref{eq:bayes}; at $b=0$ it equals
\[
\frac{\E_i[xf_{i1}(x)]-\mu_i\E_i f_{i1}(x)}{a_i}=\frac{c_i}{a_i}.
\]
Uniform continuity of $f_i'$ gives a common modulus for the quotient
derivatives, and integration supplies \eqref{eq:markexpansion} for this
entire shifted family. Thus all hypotheses of \eqref{eq:markedscore}
have been checked using the primitive assumptions.

Subtracting $\xi_i(b)$ from each cell of individual $i$ is an additive shift
and does not alter $q$. For $b=h/\sqrt{NT}$, with deterministic fixed $h$,
\eqref{eq:markedscore} and \eqref{eq:shiftedcoefficients} therefore give
\begin{align*}
    S_{NT}(b)&=-\sum_{i,t}(x_{it}-\mu_{i,b})\psi(W_{it,b})
    +\frac{\tau(1-\tau)}{2}\left\{\sum_{i=1}^{N}\frac{c_i}{a_i^2}
    +(T-1)\frac{\bar c_N}{\bar a_N^2}\right\}+\op(\sqrt{NT}) \\
    &= -\sum_{i, t} z_{it} \psi(\varepsilon_{it}) - \sum_{i, t} \ell_{it, b} 
    +\frac{\tau(1-\tau)}{2}\left\{\sum_{i=1}^{N}\frac{c_i}{a_i^2}
    +(T-1)\frac{\bar c_N}{\bar a_N^2}\right\}+\op(\sqrt{NT}),
    \numberthis \label{eq:shiftedscore}
\end{align*}
where 
\begin{align*}
    \ell_{it, b} = (x_{it}-\mu_{i,b})\psi(W_{it, b})-(x_{it}-\mu_i)\psi(\varepsilon_{it}),
\end{align*}
and $\ell_{i,b}=(x-\mu_{i,b})\psi(W_b)-(x-\mu_i)\psi(\varepsilon)$ is its value at a generic draw from
individual $i$.
Replacing the coefficients in the bias has error $o{(N+T)}=o(\sqrt{NT})$, because all
coefficient approximations are uniform and $N+T=O(\sqrt{NT})$.

The leading score increment has mean
\begin{align*}
\E[\ell_{it, b}] = \E_i[(x-\mu_{i,b})\psi(W_b)]
&=\E_i[x\{\tau-F_i(\xi_i(b)+x'b\mid x)\}]\\
&=-H_i b+O(\norm b^2).
\end{align*}
Here $\E_i[(x-\mu_i)\psi(\varepsilon)]=0$, $\E_i\psi(W_b)=0$ and $\E_i[f_{i0}(x)(x-\mu_i)]=0$.
Boundedness and the density bound imply
\begin{equation}
\norm{\ell_{i,b}}^2
\le C\ind\{|\varepsilon|\le C\norm b\}+C\norm b^2,
\qquad \sup_i\E_i\norm{\ell_{i,b}}^2\le C\norm b.
\label{eq:incrementvariance}
\end{equation}
Independence across all cells bounds the second moment of the centred
sum of these increments, divided by $\sqrt{NT}$, by
$(NT)^{-1}\sum_{i,t}\E_i\norm{\ell_{i,b}}^2\le C\norm b=O((NT)^{-1/2})$, so this sum
tends to zero in probability. The sum of the mean increments,
divided by $\sqrt{NT}$, is $-H_Nh+O((NT)^{-1/2})$.
Consequently, with
\begin{equation}
Z_{NT}=\frac1{\sqrt{NT}}\sum_{i,t}z_{it}\psi(\varepsilon_{it}),\qquad
d_{NT}^*=\frac{\tau(1-\tau)}{2\sqrt{NT}}\left\{\sum_{i=1}^{N}\frac{c_i}{a_i^2}
+(T-1)\frac{\bar c_N}{\bar a_N^2}\right\},
\label{eq:zn}
\end{equation}
we have the pointwise local subgradient expansion
\begin{equation}
\frac{S_{NT}(h/\sqrt{NT})}{\sqrt{NT}}
=-Z_{NT}+H_Nh+d_{NT}^*+\op(1).
\label{eq:localscore}
\end{equation}
Explicitly, \eqref{eq:shiftedscore} divided by $\sqrt{NT}$ reads
\[
\frac{S_{NT}(b)}{\sqrt{NT}}=-Z_{NT}-\frac1{\sqrt{NT}}\sum_{i,t}(\ell_{it,b}-\E\ell_{it,b})
-\frac1{\sqrt{NT}}\sum_{i,t}\E\ell_{it,b}+d_{NT}^*+\op(1).
\]
The centred sum is $\op(1)$, and $-(NT)^{-1/2}\sum_{i,t}\E\ell_{it,b}=H_Nh+O((NT)^{-1/2})$
by the mean increment; this is \eqref{eq:localscore}.

\subsubsection{Convexity, localisation of the slope, and its limit law}\label{sssec:convexity}
Set $G_{NT}(h)=Q_{NT}(h/\sqrt{NT})-Q_{NT}(0)$ and $t_{NT}=-Z_{NT}+d_{NT}^*$.
For fixed $h$ and any positive integer $J$, the subgradient inequality
along the line from zero to $h$ gives
\begin{equation}
\frac1J\sum_{j=0}^{J-1}h'\frac{S_{NT}(jh/(J\sqrt{NT}))}{\sqrt{NT}}
\le G_{NT}(h)\le
\frac1J\sum_{j=1}^{J}h'\frac{S_{NT}(jh/(J\sqrt{NT}))}{\sqrt{NT}}.
\label{eq:riemann}
\end{equation}
For each fixed $J$ only finitely many deterministic points occur. By
\eqref{eq:localscore}, these bounds are
$h't_{NT}+(1/2\mp1/(2J))h'H_Nh+\op(1)$.
Letting $N,T\to\infty$ and then $J\to\infty$ proves
$G_{NT}(h)-h't_{NT}\to_p h'Hh/2$ pointwise.

We recall and justify the convexity step from pointwise to uniform
convergence. Finite convex functions converging pointwise in probability
to a finite continuous function converge uniformly on compact sets in
probability. To prove this, enclose the compact set in the inner quarter
of a cube centred at zero. Convergence at the outer cube's vertices bounds the functions
above there with probability tending to one. Convergence at zero and
$F(h)\ge2F(0)-F(-h)$, for each such function $F$, bound them below on an inner cube. Convexity
then bounds their secant slopes on the innermost cube by a deterministic
constant: extend each line segment to the boundary and use the upper and
lower bounds. A finite net and continuity of the limiting function give
the uniform conclusion. For random functions, apply this along
subsequences that converge almost surely at every point of a countable
dense set, which exist by a diagonal argument (for laws that change with
$N$, after realising the arrays on a product space); this gives the
conclusion in probability. Apply this fact to the convex functions
$G_{NT}(h)-h't_{NT}$. Since $H_N\to H$, for every fixed $R$,
\begin{equation}
\sup_{\norm h\le R}\left|G_{NT}(h)-h't_{NT}-\tfrac12h'H_Nh\right|=\op(1).
\label{eq:uniformcriterion}
\end{equation}

The vectors $t_{NT}$ are tight, since $\operatorname{Var}(Z_{NT})=\tau(1-\tau) V_N$
is bounded and $d_{NT}^*$ is bounded, because $\norm{c_i}\le2C_0K$, $a_i\ge c_0$ and
$N+T=O(\sqrt{NT})$. The least eigenvalue of $H_N$ is
bounded below eventually. On a sufficiently large fixed sphere, the
quadratic in \eqref{eq:uniformcriterion} is at least $1$ with
arbitrarily high probability, since on $\norm h=R$ it is at least
$\tfrac12\lambda_{\min}(H_N)R^2-R\norm{t_{NT}}$. Since $G_{NT}(0)=0$, convexity excludes any
minimiser outside that sphere: its segment to zero would meet the sphere
at a point with $G_{NT}\le0$. Thus all slope minimisers have
$\sqrt{NT}(\widehat\beta-\beta_{0})=\Op(1)$.

The quadratic's minimiser is $h_{NT}^*=-H_N^{-1}t_{NT}$, also tight. On a
fixed ball containing both it and $\widehat h=\sqrt{NT}(\widehat\beta-
\beta_{0})$, the minimising inequality and positive curvature give
\[
\tfrac12\lambda_{\min}(H_N)\norm{\widehat h-h_{NT}^*}^2
\le2\sup_{\norm h\le R}
|G_{NT}(h)-h't_{NT}-h'H_Nh/2|.
\]
First let $N,T\to\infty$ and then let the ball radius increase. We obtain
\begin{equation}
\sqrt{NT}(\widehat\beta-\beta_{0})
=H_N^{-1}Z_{NT}-H_N^{-1}d_{NT}^*+\op(1).
\label{eq:slopelocal}
\end{equation}
Dividing by $\sqrt{NT}$ and using $(T-1)/(NT)=N^{-1}-(NT)^{-1}$
proves \eqref{eq:slope}, since the $(NT)^{-1}$ term is negligible.
Finally, $\E_i[z\psi(\varepsilon)]=0$ and
$\operatorname{Var}(Z_{NT})=\tau(1-\tau) V_N\to \tau(1-\tau) V$. The summands are uniformly
bounded before normalisation, so the central limit theorem for independent triangular arrays
(the Lindeberg--Feller theorem; \citet[Proposition~2.27]{vdv}, with the
Cram\'er--Wold device) applies. With $H_N^{-1}\to H^{-1}$, \eqref{eq:slopelocal} and Slutsky's theorem
\citep[Lemma~2.8]{vdv} then give \eqref{eq:slopeclt}. This proves Theorem~\ref{thm:slope}.

\section{Auxiliary results}\label{sec:auxiliary}

This appendix collects the deterministic quadratic identities used in
Appendix~\ref{sssec:geometry}, a general duality lemma used in
Appendix~\ref{ssec:rowscaling}, the uniform rate of the fitted effects needed
by the residual-based estimates of Sections~\ref{sec:feasible}
and~\ref{sec:spj}, and a rank argument showing that the selection rule of
Section~\ref{ssec:estimator} determines the coefficients with probability
tending to one. The fitted-effect rate is not an inference result for
the effects.

\subsection{Quadratic minimisation and trace identities}\label{ssec:quadraticgeometry}

Let $N,T\ge2$ and let
\[
\A=\left\{v\in\R^{N\times T}:v_{it}=r_i+s_t,\quad\sum_{t=1}^T s_t=0\right\}.
\]
For positive constants $a_1,\ldots,a_N$, put
\[
A=\sum_{i=1}^N a_i,\qquad \bar a=A/N,\qquad
\Lambda=\operatorname{diag}_{it}(a_i)\in\R^{NT\times NT}.
\]
Each $a_i$ appears $T$ times on the diagonal of $\Lambda$, so the weight
is constant over time within individual $i$.
The following lemmas are deterministic and hold for any such positive
weights.

\begin{lemma}[Weighted quadratic minimisation on the additive space]
\label{lem:weightedquadratic}
Let $B$ be any $NT\times(N+T-1)$ matrix whose columns form a basis of $\A$.
For every $g\in\R^{NT}$, the criterion
\[
Q_g(v)=\frac12v'\Lambda v-v'g,\qquad v\in\A,
\]
has the unique minimiser
\[
v_g=Mg,\qquad M:=B(B'\Lambda B)^{-1}B'.
\]
Here $B'\Lambda B$ is positive definite, and $M$ is symmetric,
positive semidefinite, and independent of the choice of basis.
Entrywise,
\[
(Mg)_{it}
=\frac{\bar g_{i\cdot}}{a_i}
+\frac1A\sum_{j=1}^N(g_{jt}-\bar g_{j\cdot}),
\qquad
\bar g_{i\cdot}=T^{-1}\sum_t g_{it}.
\]
Moreover, $M\Lambda M=M$ and, for every $v\in\A$,
\[
Q_g(v)
=\frac12(v-Mg)'\Lambda(v-Mg)-\frac12g'Mg.
\]
In particular, $\min_{v\in\A}Q_g(v)=-g'Mg/2$.
The matrix $M\Lambda$ is the orthogonal projection onto $\A$
for the inner product $\langle v,w\rangle_\Lambda=v'\Lambda w$.
\end{lemma}

\begin{proof}
For every nonzero $\theta\in\R^{N+T-1}$, full column rank of $B$
and positivity of the weights give
\[
\theta'B'\Lambda B\theta
=\sum_{i,t}a_i\{(B\theta)_{it}\}^2>0.
\]
Writing $v=B\theta$, the first-order condition is
$B'\Lambda B\theta=B'g$. This yields the unique minimiser $v_g=Mg$.
The matrix formula shows that $M$ is symmetric and positive semidefinite.
Since the minimiser is unique for every $g$, $M$ is independent of
the chosen basis.

For the entrywise formula, write $(Mg)_{it}=r_i+s_t$ with
$\sum_t s_t=0$. Introducing a multiplier $\lambda$ for this normalisation,
the row and time first-order conditions are
\begin{align*}
\sum_t a_i(r_i+s_t)=\sum_t g_{it},\qquad 
\sum_i a_i(r_i+s_t)+\lambda=\sum_i g_{it}.
\end{align*}
Summing the row equations over $i$ and the time equations over $t$, and subtracting, gives $T\lambda=0$.
The normalisation and the row equations therefore yield
$r_i=\bar g_{i\cdot}/a_i$. The time equations then give
\[
s_t=\frac1A\sum_j(g_{jt}-\bar g_{j\cdot}),
\]
whose sum over $t$ is zero. Adding $r_i$ and $s_t$ proves the formula.

Put $G_B=B'\Lambda B$. Direct multiplication gives
\[
M\Lambda M=BG_B^{-1}(B'\Lambda B)G_B^{-1}B'=M.
\]
The normal equations also imply
\[
w'(\Lambda v_g-g)=0\qquad\text{for every }w\in\A.
\]
Because $v-v_g\in\A$, expansion around $v_g$ yields
\[
Q_g(v)=Q_g(v_g)+\frac12(v-v_g)'\Lambda(v-v_g).
\]
Symmetry and $M\Lambda M=M$ give $v_g'\Lambda v_g=g'Mg$,
and hence $Q_g(v_g)=-g'Mg/2$.

Finally, for any $y\in\R^{NT}$, the objective with $g=\Lambda y$
differs from $(v-y)'\Lambda(v-y)/2$ by a constant independent of $v$.
Thus its minimiser $M\Lambda y$ is the $\Lambda$-orthogonal projection
of $y$ onto $\A$.
\end{proof}

\begin{lemma}[Trace identities for additive quadratic minimisation]\label{lem:weightedtraces}
Under the notation and assumptions of Lemma~\ref{lem:weightedquadratic},
\[
\operatorname{tr}M=\sum_{i=1}^N\frac1{a_i}+\frac{T-1}{\bar a},
\qquad
\operatorname{tr}(M^2)=\sum_{i=1}^N\frac1{a_i^2}+\frac{T-1}{\bar a^2}.
\]
If $\min_i a_i\ge c>0$, then
\[
\operatorname{tr}M\le\frac{N+T-1}{c},\qquad
\operatorname{tr}(M^2)\le\frac{N+T-1}{c^2}.
\]
\end{lemma}

\begin{proof}
Use the basis $B=[R\ C]$, where the individual indicators and common
time contrasts are
\begin{align*}
R^{(j)}_{it}&=\ind\{i=j\},\qquad j=1,\ldots,N,\\
C^{(\ell)}_{it}&=\ind\{t=\ell\}-\ind\{t=T\},\qquad \ell=1,\ldots,T-1.
\end{align*}
The two sets of columns span the individual-constant and common
time-contrast subspaces of $\A$. Since $\sum_t s_t=0$ for a time
contrast and $a_i$ is constant over time, these subspaces are orthogonal
under both the Euclidean and $\Lambda$-weighted inner products.

Put $D=\operatorname{diag}(a_1,\ldots,a_N)$ and $G=C'C$.
The columns of $C$ are linearly independent, so $G$ is positive definite.
Because each time contrast is repeated for every individual,
\[
B'B=
\begin{pmatrix}
T I_N&0\\
0&G
\end{pmatrix},
\qquad
B'\Lambda B=
\begin{pmatrix}
T D&0\\
0&\bar a G
\end{pmatrix}.
\]
Consequently,
\[
K:=(B'\Lambda B)^{-1}B'B
=
\begin{pmatrix}
D^{-1}&0\\
0&\bar a^{-1}I_{T-1}
\end{pmatrix}.
\]
Writing $G_B=B'\Lambda B$, the cyclic property of the trace gives
\begin{align*}
\operatorname{tr}M
&=\operatorname{tr}(G_B^{-1}B'B)=\operatorname{tr}K,\\
\operatorname{tr}(M^2)
&=\operatorname{tr}(G_B^{-1}B'B\,G_B^{-1}B'B)
=\operatorname{tr}(K^2).
\end{align*}
Taking the traces of the block-diagonal matrices proves the closed
forms. If $a_i\ge c$ for every $i$, then $\bar a\ge c$ as well, which
gives the stated bounds.
\end{proof}

\subsection{Quantile-loss duality on a linear subspace}\label{ssec:quantileduality}

\begin{lemma}[Quantile-loss duality]\label{lem:quantileduality}
Fix $0<\tau<1$ and a linear subspace $\mathcal V\subseteq\R^{N\times T}$, and define
\[
\D_{\mathcal V}
=\{q\in\mathcal V^\perp:\tau-1\le q_{it}\le\tau\text{ for all }i,t\}.
\]
For every $w\in\R^{N\times T}$,
\[
\min_{v\in\mathcal V}\sum_{i,t}\rho(w_{it}-v_{it})
=\max_{q\in\D_{\mathcal V}}w'q,
\]
with both optima attained. Every optimal pair $(\widehat v,\widehat q)$ satisfies
\[
\widehat v'\widehat q=0,\qquad
\widehat q_{it}\in\partial\rho(w_{it}-\widehat v_{it}).
\]
Moreover, there is a Borel maximising selection $q_{\mathcal V}(w)$ such that
$q_{\mathcal V}(w+a)=q_{\mathcal V}(w)$ for every $a\in\mathcal V$.
\end{lemma}

\begin{proof}
Let $B$ be a basis matrix of $\mathcal V$ and write $v=B\theta$.
Since $\rho(e)=\max\{\tau e,(\tau-1)e\}$, the primal problem is equivalent to the linear programming problem:
\[
\begin{aligned}
\min_{\theta,\ell}\quad&\mathbf1'\ell\\
\text{subject to}\quad&
\ell+\tau B\theta\ge\tau w,\\
&\ell+(\tau-1)B\theta\ge(\tau-1)w,
\end{aligned}
\]
where $\theta$ and $\ell$ are unrestricted in sign and inequalities hold componentwise.
For fixed $\theta$, minimising over $\ell$ sets
$\ell_{it}=\rho(w_{it}-(B\theta)_{it})$.
Its standard LP dual is
\[
\begin{aligned}
\max_{\lambda^+,\lambda^-}\quad&
w'\{\tau\lambda^++(\tau-1)\lambda^-\}\\
\text{subject to}\quad&
\lambda^++\lambda^-=\mathbf1,\\
&B'\{\tau\lambda^++(\tau-1)\lambda^-\}=0,\\
&\lambda^+,\lambda^-\ge0.
\end{aligned}
\]
The transformation $q=\tau\lambda^++(\tau-1)\lambda^-$, with inverse
$\lambda^+=q-(\tau-1)\mathbf1$ and $\lambda^-=\tau\mathbf1-q$,
identifies these constraints with $q\in\D_{\mathcal V}$, since
$B'q=0$ is equivalent to $q\in\mathcal V^\perp$.
The primal LP is feasible, for example at $\theta=0$ and
$\ell_{it}=\rho(w_{it})+1$, and its objective is bounded below by zero.
Strong duality of linear programming \citep[Section~5.2.4, p.~227]{boydvandenberghe2004} therefore gives the equality of values and attainment of both optima.

For every feasible pair $(v,q)$, orthogonality gives $v'q=0$.
Writing $e_{it}=w_{it}-v_{it}$, its primal--dual gap is
\[
\sum_{i,t}\rho(e_{it})-w'q
=\sum_{i,t}\{\rho(e_{it})-e_{it}q_{it}\}\ge0.
\]
Each summand is nonnegative because $q_{it}\in[\tau-1,\tau]$.
At an optimal pair the gap is zero, so equality holds cellwise.
This is equivalent to $q_{it}\in\partial\rho(e_{it})$, proving complementary slackness.

Finally, $\D_{\mathcal V}$ is a nonempty compact polytope containing $0$.
Choose $q_{\mathcal V}(w)$ as the first maximising vertex in a fixed enumeration of its vertices.
This choice is Borel because it is determined by finitely many comparisons of linear functions of $w$.
For $a\in\mathcal V$ and $q\in\D_{\mathcal V}$, $(w+a)'q=w'q$;
hence the selected vertex is unchanged by such shifts.
\end{proof}

\subsection{Uniform rate of the fitted effects}

\begin{lemma}[Uniform rate of the fitted effects]\label{lem:uniformrate}
Under A1--A4,
\begin{equation}
\max_i|\widehat\alpha_{i}-\alpha_{i0}|
+\max_t|\widehat\gamma_{t}-\gamma_{t0}|=\Op(\min(N,T)^{-1/3}).
\label{eq:maxrate}
\end{equation}
\end{lemma}

\begin{proof}
Write $\widehat b=\widehat\beta-\beta_{0}$ and
$\widehat v_{it}=(\widehat\alpha_{i}-\alpha_{i0})+
(\widehat\gamma_{t}-\gamma_{t0})$. At the selected slope, $\widehat v$
minimises the additive criterion for the array $\varepsilon-X\widehat b$. Since $\widehat\gamma$ and
$\gamma_0$ both sum to zero, the individual and time components of $\widehat v$ in the normalisation of
Appendix~\ref{ssec:additive} are $\widehat\alpha_{i}-\alpha_{i0}$ and $\widehat\gamma_{t}-\gamma_{t0}$.
Theorem~\ref{thm:slope} gives
$\norm{X\widehat b}_\infty\le K\norm{\widehat b}=\Op((NT)^{-1/2})$.
Apply Lemma~\ref{lem:additive} with $W=\varepsilon$, whose individual laws satisfy
\eqref{eq:scalarconditions} by Appendix~\ref{sssec:localslope} with $b=0$, and
$u=-X\widehat b$; its simultaneous bound over perturbation balls permits
this random perturbation. Since the lemma is stated for a fixed $L$,
given $\epsilon>0$ choose $L_\epsilon$ with
$\Pp(\sqrt{NT}K\norm{\widehat b}>L_\epsilon)<\epsilon$ for large $N,T$, by
Theorem~\ref{thm:slope}, and apply the lemma with $d_{NT}=L_\epsilon/\sqrt{NT}$
outside this event. The maximum-norm bound
\eqref{eq:nuisancemax} proves \eqref{eq:maxrate}.
\end{proof}

\subsection{Rank and equivariance}\label{ssec:rank}
The fitted-array space of the joint regression is finite-dimensional and
closed. Quantile loss is coercive in the fitted array, and its minimiser set is
compact and convex. Squared residual loss is strictly convex in that
array, giving a unique measurable secondary minimiser equivariant under
the transformations of Section~\ref{ssec:estimator}.

To check rank, let $\Xi_{it}=x_{it}-\E_i x$ and
$\Sigma_N=N^{-1}\sum_{i=1}^{N}\operatorname{Var}_i(x)$.
The weighted-variance minimising property of $\mu_i$ gives
$H_i\preceq C\operatorname{Var}_i(x)$, hence
$\Sigma_N\succeq H_N/C$ is uniformly positive definite eventually.
Independence and boundedness give
$(NT)^{-1}\Xi'\Xi-\Sigma_N\to_p0$. Also
$\E\operatorname{tr}(\Xi'P\Xi)\le C{(N+T)}$, so
$(NT)^{-1}\Xi'P\Xi\to_p0$ by Markov's inequality. The array of individual means
$\E_i x$ belongs to $\A$, and therefore
\begin{equation}
\frac1{NT}X'(I-P)X-\Sigma_N\longrightarrow_p0.
\label{eq:rank}
\end{equation}
Thus the design is full rank modulo the one normalisation with probability
tending to one. Write $E_{NT}$ for this event; it is invariant under individual and time permutations. On $E_{NT}$, the secondary fitted-array rule determines
unique coefficients. Any convention on $E_{NT}^c$ contributes only $o(1)$
to the probability statements below.

\section{Proof of Theorem~\ref{thm:feasible}}\label{sec:prooffeasible}

\begin{proof}
Constants $C$ below depend only on the constants in A1--A3, on $K_0$, and on $\underline c$ and $\bar c$, and may
change from line to line.
Theorem~\ref{thm:slope} and Lemma~\ref{lem:uniformrate} imply
\[
\delta_{NT}:=\max_{i,t}|\widehat\varepsilon_{it}-\varepsilon_{it}|
=\Op(\min(N,T)^{-1/3}).
\]
For $u\ge0$ let $G_i(u)=\Pp_i(|\varepsilon|\le u)$ and
$G_{i,T}(u)=T^{-1}\sum_{t=1}^{T}\ind\{|\varepsilon_{it}|\le u\}$, and let both be zero for $u<0$. With
$r_1=\min\{r_0,c_0/(2C_0)\}$ as in Appendix~\ref{sssec:localslope}, $c_0/2\le f_i(u\mid x)\le C_0$ for $|u|\le r_1$, so that
\[
c_0u\le G_i(u)\le2C_0u\qquad(0\le u\le r_1).
\]
The absolute errors of an individual are iid. As for the individuals in \eqref{eq:cdf}, the
Dvoretzky--Kiefer--Wolfowitz inequality (\citet{dkw}; \citet{massart}) and a union bound show that the event
\[
\mathcal E_{NT}=\Bigl\{\max_i\sup_{u}|G_{i,T}(u)-G_i(u)|\le\vartheta_T\Bigr\},
\qquad \vartheta_T=\sqrt{\log N/T},
\]
has probability at least $1-2/N$; under A4, $\vartheta_T=o(\min(N,T)^{-1/3})$. Finally, since
$\bigl||\widehat\varepsilon_{it}|-|\varepsilon_{it}|\bigr|\le\delta_{NT}$, for every $i$ and $u\ge0$,
\[
TG_{i,T}(u-\delta_{NT})\le\#\{t:|\widehat\varepsilon_{it}|\le u\}\le TG_{i,T}(u+\delta_{NT}).
\]

\emph{Step 1: the nearest-neighbour bandwidths.} Let $\pi$ be $\pi_a$ or $\pi_c$. By the second line of equation 
\eqref{eq:bandwidth} in Assumption~A5, $\pi\to0$ and $\min(N,T)^{1/3}\pi\to\infty$ (for $\pi_c$ because $\pi_c\ge\pi_c^2$ once $\pi_c\le1$), so that,
with probability tending to one, $\mathcal E_{NT}$ holds, $2\pi/c_0\le r_1$, and
$(2C_0+c_0)\delta_{NT}+\vartheta_T+T^{-1}<\pi/2$. On this event $\delta_{NT}<\pi/(4C_0)\le2\pi/c_0$, so that the
arguments of $G_i$ below lie in $[0,r_1]$, and for every $i$,
\[
\#\{t:|\widehat\varepsilon_{it}|\le2\pi/c_0\}\ge T\{G_i(2\pi/c_0-\delta_{NT})-\vartheta_T\}
\ge T(2\pi-c_0\delta_{NT}-\vartheta_T)\ge\pi T+1\ge\lceil\pi T\rceil,
\]
so that $\widehat h_i(\pi)\le2\pi/c_0$, and
\[
\#\{t:|\widehat\varepsilon_{it}|\le\pi/(4C_0)\}\le T\{G_i(\pi/(4C_0)+\delta_{NT})+\vartheta_T\}
\le T(\pi/2+2C_0\delta_{NT}+\vartheta_T)<\pi T\le\lceil\pi T\rceil,
\]
so that $\widehat h_i(\pi)>\pi/(4C_0)$. This is the first claim; the fractions $T^{-1/5}$ and $T^{-1/7}$ satisfy
the second line of \eqref{eq:bandwidth} in A5 because $T\asymp\min(N,T)$ under A4.

\emph{Step 2: true errors, uniformly over the bandwidths.} For a scalar mark $q(x)$ with $|q|\le1$ and
$h>0$, put
\[
A_i(h)=\frac1{Th}\sum_{t=1}^{T} q(x_{it})K_0(\varepsilon_{it}/h),\qquad
\dot A_i(h)=-\frac1{Th^2}\sum_{t=1}^{T} q(x_{it})K_0'(\varepsilon_{it}/h),
\]
and let $\mathcal I_a=[\underline c\,\pi_a,\bar c\,\pi_a]$ and $\mathcal I_c=[\underline c\,\pi_c,\bar c\,\pi_c]$. We
show that
\begin{equation}
\max_i\sup_{h\in\mathcal I_a}\bigl|A_i(h)-\E_i[q(x)f_{i0}(x)]\bigr|=\op(1),\qquad
\max_i\sup_{h\in\mathcal I_c}\bigl|\dot A_i(h)-\E_i[q(x)f_{i1}(x)]\bigr|=\op(1).
\label{eq:kernelbounds}
\end{equation}
Take $N$ and $T$ so large that $2\bar c(\pi_a\vee\pi_c)\le r_1\wedge1$ and
$\vartheta_T\le\underline c(\pi_a\wedge\pi_c)$; all bandwidths below are then at most $r_1\wedge1$ and at least
$\vartheta_T$. For such $h$, a change of variables gives
$\E A_i(h)=\int\E_i[q(x)f_i(hu\mid x)]K_0(u)\,du$, which differs from
$\E_i[q(x)f_{i0}(x)]$ by at most $C_0h$ because $|f_i'|\le C_0$. For the derivative, change variables and
integrate by parts:
\[
\E\dot A_i(h)=-\frac1h\int\E_i[q(x)f_i(hu\mid x)]K_0'(u)\,du
=\int\E_i[q(x)f_i'(hu\mid x)]K_0(u)\,du.
\]
Compact support removes the boundary term, and the common continuity modulus in A2 bounds the difference from
$\E_i[q(x)f_{i1}(x)]$ by $\omega_0(h)$. Both bounds are uniform in $i$ and vanish uniformly over the two intervals.

Next fix a bandwidth. Since $\Pp_i(|\varepsilon|\le h\mid x)\le2C_0h$, the summands of $A_i(h)$ are bounded by
$C/h$ and have variance at most $C/h$, and those of $\dot A_i(h)$ are bounded by $C/h^2$ and have variance at
most $C/h^3$. Bernstein's inequality (\citet[Lemma~2.2.10 and Example~2.2.11]{vdvw}) gives, for $0<s\le1$,
\[
\Pp\{|A_i(h)-\E A_i(h)|>s\}\le2\exp(-Ths^2/C),\qquad
\Pp\{|\dot A_i(h)-\E\dot A_i(h)|>s\}\le2\exp(-Th^3s^2/C).
\]
Let $\eta=\pi_c^2$, and cover $\mathcal I_a$ and $\mathcal I_c$ by the grids
$h^{(j)}=\underline c\,\pi_a(1+\eta)^j$ and $g^{(j)}=\underline c\,\pi_c(1+\eta)^j$, $j=0,\ldots,J$, where $J$ is
the least integer with $(1+\eta)^J\ge\bar c/\underline c$, so that $J\le C/\eta$. A union bound over the individuals and
the grid points bounds the probability that a centred average at a grid point exceeds $s$ by
\[
2N(J+1)\bigl\{\exp(-\underline c\,T\pi_as^2/C)+\exp(-\underline c^3T\pi_c^3s^2/C)\bigr\}.
\]
This tends to zero for each fixed $s$: by Assumption~A5, $T\pi_a\ge T\min(N,T)^{-1/3}$,
$T\pi_c^3\ge T\min(N,T)^{-1/2}$ and $J\le C\min(N,T)^{1/3}$ for large $N$ and $T$, and
$N\asymp T\asymp\min(N,T)$ under A4. Between grid points, differentiate in the bandwidth. Because $K_0$, $K_0'$ and
$K_0''$ vanish outside $[-1,1]$,
\[
\Bigl|\frac{\partial}{\partial h}\Bigl\{\frac{K_0(e/h)}{h}\Bigr\}\Bigr|\le\frac{C}{h^2}\ind\{|e|\le h\},\qquad
\Bigl|\frac{\partial}{\partial h}\Bigl\{\frac{K_0'(e/h)}{h^2}\Bigr\}\Bigr|\le\frac{C}{h^3}\ind\{|e|\le h\}.
\]
Hence, on $\mathcal E_{NT}$, for $h^{(j)}\le h\le(1+\eta)h^{(j)}$ and for $g^{(j)}\le g\le(1+\eta)g^{(j)}$,
\begin{align*}
|A_i(h)-A_i(h^{(j)})|&\le\frac{C\eta}{h^{(j)}}\,G_{i,T}\{(1+\eta)h^{(j)}\}\le C\eta,\\
|\dot A_i(g)-\dot A_i(g^{(j)})|&\le\frac{C\eta}{(g^{(j)})^2}\,G_{i,T}\{(1+\eta)g^{(j)}\}
\le\frac{C\eta}{g^{(j)}}\le C\pi_c,
\end{align*}
where we use $G_{i,T}(u)\le2C_0u+\vartheta_T\le Cu$ for $\vartheta_T\le u\le r_1$. With the bounds on the
expectations, this proves \eqref{eq:kernelbounds}.

\emph{Step 3: residuals.} Let $\widehat A_i(h)$ and $\widehat{\dot A}_i(h)$ be $A_i(h)$ and $\dot A_i(h)$ with
$\widehat\varepsilon_{it}$ in place of $\varepsilon_{it}$. If $|\varepsilon_{it}|>h+\delta_{NT}$, then
$|\widehat\varepsilon_{it}|>h$, and the kernel and its derivative vanish at both arguments. Otherwise the mean
value theorem bounds the changes in $K_0$ and in $K_0'$ by $C\delta_{NT}/h$. Hence, on $\mathcal E_{NT}$, for
every $i$ and every $h\le r_1/2$ with $\delta_{NT}\vee\vartheta_T\le h$,
\[
|\widehat A_i(h)-A_i(h)|\le\frac{C\delta_{NT}}{h^2}\,G_{i,T}(h+\delta_{NT})\le\frac{C\delta_{NT}}{h},\qquad
|\widehat{\dot A}_i(h)-\dot A_i(h)|\le\frac{C\delta_{NT}}{h^3}\,G_{i,T}(h+\delta_{NT})\le\frac{C\delta_{NT}}{h^2}.
\]
These bounds hold for all such bandwidths at once, and so at bandwidths that depend on the data. With
probability tending to one, every $h$ in $\mathcal I_a\cup\mathcal I_c$ satisfies the two restrictions, because
$\delta_{NT}=\op(\pi_a\wedge\pi_c)$; over $\mathcal I_a$ and $\mathcal I_c$ the bounds are then
$\Op(\min(N,T)^{-1/3}/\pi_a)$ and $\Op(\min(N,T)^{-1/3}/\pi_c^2)$, which tend to zero by Assumption~A5.
With \eqref{eq:kernelbounds} and the first line of equation \eqref{eq:bandwidth} in A5,
\[
\max_i\bigl|\widehat A_i(h_i)-\E_i[q(x)f_{i0}(x)]\bigr|=\op(1),\qquad
\max_i\bigl|\widehat{\dot A}_i(g_i)-\E_i[q(x)f_{i1}(x)]\bigr|=\op(1).
\]

\emph{Step 4: the estimators.} Apply these conclusions to $q=1$ and, after division by $K$ or $K^2$, to each
coordinate of $x$ and each entry of $xx'$. First, $\widehat a_i$ is $\widehat A_i(h_i)$ with $q=1$. It is
uniformly consistent for $a_i\ge c_0$, so that $\min_i\widehat a_i\ge c_0/2$ with probability tending to one,
and division yields uniform consistency of $\widehat\mu_i$. Next, $\widehat c_i$ is $\widehat{\dot A}_i(g_i)$ with
$q=x$ minus $\widehat\mu_i$ times $\widehat{\dot A}_i(g_i)$ with $q=1$, and
$\E_i[f_{i1}(x)x]-\mu_i\E_i f_{i1}(x)=c_i$. This gives \eqref{eq:uniformdensityconsistency}. The term of
$\widehat H$ for individual $i$ is $\widehat A_i(h_i)$ with $q=xx'$ minus
$\widehat a_i\widehat\mu_i\widehat\mu_i'$, and $\E_i[f_{i0}(x)xx']-a_i\mu_i\mu_i'=H_i$, so that
$\widehat H-H_N\to_p0$.
For $\widehat V$, uniform individual LLNs for bounded first and second moments
(by Hoeffding's inequality and a union bound over the individuals), combined with uniform consistency of $\widehat\mu_i$, give the claim.

All averages and inverses in \eqref{eq:biasestimators} are continuous
on uniformly positive density and matrix neighbourhoods, proving their
consistency. Since $\sqrt{NT}/T$ and $\sqrt{NT}/N$ are bounded,
plugging these consistent bias coefficients into \eqref{eq:slopeclt}
proves \eqref{eq:abcclt}. Slutsky's theorem proves the studentised
statement. The studentised statistic needs
$\ell'\Omega\ell>0$: since $f_{i0}\le C_0$, $H_i\preceq C_0V_i$, so that
$V\succeq H/C_0\succ0$ and $\Omega\succ0$.
\end{proof}

\section{Proof of Theorem~\ref{thm:spj}}\label{sec:proofspj}
\begin{proof}
All individual population moments used in $H_N,V_N,\bar a_N,B_N,D_N$ are
uniformly bounded. For a simple random sample of a fixed positive fraction of the individuals, the conditional-on-population variance of each sample mean
is $O(N^{-1})$. Hence each individual half's moments differ from the full
panel's by $O_p(N^{-1/2})$. Positivity of $H_N$ and $\bar a_N$ then
gives the same rate for the half-panel matrices, inverses, and bias
coefficients. The primitive probability bounds proved above are uniform,
so the slope theorem applies conditionally on the split and hence
unconditionally. In detail, let $\mathcal G_N$ be the set of splits whose
half-panel moments are within $\eta_N$ of the full panel's, with $\eta_N\to0$
and $\eta_NN^{1/2}\to\infty$, so that $\Pp(S\notin\mathcal G_N)\to0$. Since the
split $S$ is independent of the data, for the event $A$ that a half-panel
remainder exceeds $\epsilon(NT)^{-1/2}$,
\[
\Pp(A)=\sum_s\Pp(S=s)\Pp(A\mid S=s)
\le\Pp(S\notin\mathcal G_N)+\sup_{s\in\mathcal G_N}\Pp(A\mid S=s).
\]
The supremum tends to zero: otherwise there would be splits
$s_N\in\mathcal G_N$ along which it stays above some $\epsilon'>0$; along a
further subsequence $N_j/N$ and the half-panel moments converge, so that A1--A4 hold
for the half panels given $S=s_N$, and Theorem~\ref{thm:slope} gives a
contradiction.

Let $L_{NT}=H_N^{-1}(NT)^{-1}\sum_{i,t}z_{it}\psi(\varepsilon_{it})$.
The time halves have the same population moments, and satisfy A1--A4 along subsequences on which $T_j/T$
converges, which suffices for statements in probability. In each half,
recentre the true time effects to sum to zero and shift the individual
effects to compensate, which changes neither the data nor the slope.
Size-weighting their expansions then gives
\[
\bar\beta_T-\beta_{0}
=L_{NT}+2B_N/T+D_N/N+\op((NT)^{-1/2}).
\]
For individual halves, replacing each half's $H_{\mathcal R_j}^{-1}$ by $H_N^{-1}$ in
its weighted score costs $\op((NT)^{-1/2})$. Conditional on the split,
the score of each half is $O_p(\sqrt{NT})$ before normalisation, while
the matrix difference is $O_p(N^{-1/2})$. Their weighted leading score
is consequently $L_{NT}+\op((NT)^{-1/2})$. Size-weighting the biases gives
\[
\bar\beta_N-\beta_{0}
=L_{NT}+B_N/T+2D_N/N+\op((NT)^{-1/2}).
\]
For example, the weighted time-effect bias is
$\sum_{j=1}^2(N_j/N)D_{\mathcal R_j}/N_j=(D_{\mathcal R_1}+D_{\mathcal R_2})/N
=2D_N/N+\op((NT)^{-1/2})$, where $D_{\mathcal R_j}$ is the coefficient $D$ of the $j$th individual half.
Every half-panel remainder has order $\op((NT)^{-1/2})$ because its sample
size is a fixed fraction of $NT$. Substitution into \eqref{eq:spj}
cancels both bias terms and retains one copy of $L_{NT}$. Since $\sqrt{NT}L_{NT}\Rightarrow\mathcal N(0,\Omega)$,
as in the proof of Theorem~\ref{thm:slope}, and an $\op(1)$ term does not change the limit law
\citep[Theorem~2.7(iv)]{vdv}, this proves
\eqref{eq:spjclt}.
\end{proof}

\section{Proof of Theorem~\ref{thm:spjvariance}}\label{sec:proofvariance}

\begin{proof}
\emph{Complementary slackness.} For a fit $\mathcal P$ with regressor array $X$ and dual set
$\mathcal D_{\mathcal P}=\{q:\text{zero individual and time sums},\ \tau-1\le q\le\tau\}$, linear-programming duality, as
in~\eqref{eq:dual}, gives $\min\sum\rho(Y-X\beta-v)=\max\{Y'q:q\in\mathcal D_{\mathcal P},X'q=0\}$ with both optima
attained. For a primal minimiser and a dual maximiser, $Y'\widehat q=\widehat\varepsilon'\widehat q$ because
$X'\widehat q=0$ and the fitted additive array is orthogonal to $\widehat q$. For $q\in[\tau-1,\tau]$,
$\rho(u)-uq\ge0$, with equality if and only if $q\in\partial\rho(u)$. Since
$\sum\{\rho(\widehat\varepsilon_{it})-\widehat\varepsilon_{it}\widehat q_{it}\}=0$, every term vanishes, so
$\widehat q^{\,\mathcal P}_{it}\in\partial\rho(\widehat\varepsilon^{\,\mathcal P}_{it})$ for the selected fit,
and $\widehat q^{\,\mathcal P}_{it}=\psi(\widehat\varepsilon^{\,\mathcal P}_{it})$ whenever
$\widehat\varepsilon^{\,\mathcal P}_{it}\ne0$. Hence $\widehat q^{\,\mathcal P}_{it}\ne\psi_{it}:=\psi(\varepsilon_{it})$
only if $\widehat\varepsilon^{\,\mathcal P}_{it}=0$ or $\widehat\varepsilon^{\,\mathcal P}_{it}$ and $\varepsilon_{it}$
lie on opposite sides of zero; in each case
$|\varepsilon_{it}|\le\delta_{\mathcal P}:=\max_{(i,t)\in\mathcal P}|\widehat\varepsilon^{\,\mathcal P}_{it}-\varepsilon_{it}|$.

\emph{Uniform residual errors.} The proof of Theorem~\ref{thm:feasible} shows that
$\delta_{\mathcal F}=\Op(\min(N,T)^{-1/3})$, from Theorem~\ref{thm:slope} and Lemma~\ref{lem:uniformrate}. Each time half satisfies
A1--A4 with its own number of periods (along subsequences on which $T_j/T$ converges, which suffices for conclusions
in probability) and the same individuals and population moments; recentring its true time effects changes neither the
data nor the residuals. So $\delta_{\mathcal T_j}\to_p0$. For an individual half, let $\mathcal G_N$ be the set of splits
whose half-panel matrices $H$, $V$ and $\bar a$ are within $\eta_N$ of the full panel's, as in the proof of
Theorem~\ref{thm:spj}, and fix $\epsilon>0$. Then
$\Pp(\delta_{\mathcal R_j}>\epsilon)\le\Pp(S\notin\mathcal G_N)+\sup_{s\in\mathcal G_N}\Pp(\delta_{\mathcal R_j}>\epsilon\mid S=s)$.
If the supremum did not tend to zero, there would be splits $s_N\in\mathcal G_N$ along which it stays above some
$\epsilon'>0$; along a further subsequence $N_j/N$ converges, and given $S=s_N$ the half is a deterministic
sub-panel whose moments converge to those of the full panel, so that A1--A4 hold for it and
Theorem~\ref{thm:slope} and Lemma~\ref{lem:uniformrate} give a contradiction. So $\delta_{\mathcal P}\to_p0$ for every
$\mathcal P$.

\emph{Shares of changed dual values.} Fix $\delta\in(0,r_0]$. On $\{\delta_{\mathcal P}\le\delta\}$,
\[
\frac1{N_{\mathcal P}T_{\mathcal P}}\sum_{(i,t)\in\mathcal P}|\widehat q^{\,\mathcal P}_{it}-\psi_{it}|^2
\le\frac1{N_{\mathcal P}T_{\mathcal P}}\sum_{(i,t)\in\mathcal P}\ind\{|\varepsilon_{it}|\le\delta\},
\]
because $|\widehat q^{\,\mathcal P}_{it}-\psi_{it}|\le1$. The right side is an average of independent indicators
with means at most $2C_0\delta$ by A2, and variance at most $(4N_{\mathcal P}T_{\mathcal P})^{-1}$, uniformly over splits, so it is
at most $2C_0\delta+\op(1)$. Given $\epsilon>0$, choose $\delta$ with $2C_0\delta<\epsilon$: the left side is below
$\epsilon$ with probability tending to one. Hence
$\varsigma_{\mathcal P}:=(N_{\mathcal P}T_{\mathcal P})^{-1}\sum_{(i,t)\in\mathcal P}|\widehat q^{\,\mathcal P}_{it}-\psi_{it}|^2\to_p0$.

\emph{Conclusion.} Write $\iota_{it}=H_N^{-1}z_{it}\psi_{it}$ and
$\widehat d_{it}=3(\widehat q^{\,\mathcal F}_{it}-\psi_{it})-(\widehat q^{\,\mathcal T(t)}_{it}-\psi_{it})
-(\widehat q^{\,\mathcal R(i)}_{it}-\psi_{it})$, so that, since $3-1-1=1$,
\[
\widehat\phi_{it}-\iota_{it}=(\widehat H^{-1}\widehat z_{it}-H_N^{-1}z_{it})(\psi_{it}+\widehat d_{it})
+H_N^{-1}z_{it}\widehat d_{it}.
\]
Here $|\psi_{it}+\widehat d_{it}|\le5$; $\norm{z_{it}}\le2K$, because $\norm x\le K$ by A3 and $\mu_i$ is a
weighted mean of $x$; and $\max_{i,t}\norm{\widehat H^{-1}\widehat z_{it}-H_N^{-1}z_{it}}\to_p0$ by
Theorem~\ref{thm:feasible}, since $\widehat H-H_N\to_p0$, $H_N\to H\succ0$ and
$\max_i\norm{\widehat\mu_i-\mu_i}\to_p0$. By $(a+b+c)^2\le3(a^2+b^2+c^2)$, and because every cell lies in one time
half and one individual half with $N_{\mathcal P}T_{\mathcal P}\le NT$, $(NT)^{-1}\sum_{i,t}\widehat d_{it}^{\,2}\le3(9\varsigma_{\mathcal F}
+\varsigma_{\mathcal T_1}+\varsigma_{\mathcal T_2}+\varsigma_{\mathcal R_1}+\varsigma_{\mathcal R_2})\to_p0$. Hence
$(NT)^{-1}\sum_{i,t}\norm{\widehat\phi_{it}-\iota_{it}}^2\to_p0$. As $(NT)^{-1}\sum_{i,t}\norm{\iota_{it}}^2$ is bounded,
the Cauchy--Schwarz inequality gives $NT\,\widehat V_J-(NT)^{-1}\sum_{i,t}\iota_{it}\iota_{it}'\to_p0$. Finally,
$(NT)^{-1}\sum_{i,t}\iota_{it}\iota_{it}'=H_N^{-1}\{(NT)^{-1}\sum_{i,t}z_{it}z_{it}'\psi_{it}^2\}H_N^{-1}$, whose summands
are bounded and independent, with $\E_i[z_{it}z_{it}'\psi_{it}^2]=\tau(1-\tau) V_i$ because
$\E\{\psi(\varepsilon)^2\mid x\}=\tau^2(1-\tau)+(1-\tau)^2\tau=\tau(1-\tau)$ by A2. Chebyshev's inequality gives
$NT\,\widehat V_J\to_p\tau(1-\tau) H^{-1}VH^{-1}=\Omega$.

\emph{Size.} By the first part, $\Omega\succ0$ (proof of Theorem~\ref{thm:feasible}) and continuity of
$A\mapsto A^{-1/2}$ on positive-definite matrices, $(NT\,\widehat V_J)^{-1/2}\to_p\Omega^{-1/2}$. With
$\sqrt{NT}(\widehat\beta_{\mathrm{SPJ}}-\beta_{0})\Rightarrow\mathcal N(0,\Omega)$ from Theorem~\ref{thm:spj}, Slutsky's
theorem gives $\widehat V_J^{-1/2}(\widehat\beta_{\mathrm{SPJ}}-\beta_{0})\Rightarrow\mathcal N(0,I_p)$. Under the null,
the Wald statistic is the squared norm of this vector, with limit $\chi^2_p$, and
$(\ell'\widehat\beta_{\mathrm{SPJ}}-r^\dagger)/(\ell'\widehat V_J\ell)^{1/2}
=\sqrt{NT}\,\ell'(\widehat\beta_{\mathrm{SPJ}}-\beta_{0})/(NT\,\ell'\widehat V_J\ell)^{1/2}$ has limit $\mathcal N(0,1)$
because $\ell'\Omega\ell>0$. Both limit laws are
continuous, so the null rejection probabilities converge to $\varpi$.

\emph{Uniformity.} Suppose that the convergence is not uniform. Then there are $\epsilon>0$, a subsequence of panel
sizes, and, along it, effect arrays and arrays of individual laws in the stated class under which a null rejection
probability differs from $\varpi$ by more than $\epsilon$. Since $f_i(0\mid x)\ge c_0$, $H_i\succeq c_0V_i$, so that
$H_N\succeq c_0v_0I_p$; also $\bar a_N\ge c_0$; and $H_N$, $V_N$ and $\bar a_N$ are bounded by A2 and $\norm x\le K$.
Along a further subsequence they converge, to limits with $H\succ0$, $V\succeq v_0I_p$ and $\bar a>0$, so that A1--A4
hold along it. The results used above hold for individual laws that change with $N$ (Section~\ref{sec:model}), as the proofs of
Theorem~\ref{thm:spj} and of the first part already use for the individual halves, and for arbitrary deterministic
effects. The first part of the
proof therefore gives null rejection probabilities converging to $\varpi$ along the subsequence, a contradiction.
\end{proof}


\newpage

\begin{thebibliography}{39}
\newcommand{\enquote}[1]{``#1''}
\expandafter\ifx\csname natexlab\endcsname\relax\def\natexlab#1{#1}\fi

\bibitem[\protect\citeauthoryear{Abowd, Kramarz, and Margolis}{Abowd
  et~al.}{1999}]{AbowdKramarzMargolis99}
\textsc{Abowd, J.~M., F.~Kramarz, and D.~N. Margolis} (1999): \enquote{High
  Wage Workers and High Wage Firms,} \emph{Econometrica}, 67, 251--333.

\bibitem[\protect\citeauthoryear{Andrews, Cheng, and Guggenberger}{Andrews et~al.}{2020}]{acg}
\textsc{Andrews, D. W.~K., X.~Cheng, and P.~Guggenberger} (2020): \enquote{Generic Results for Establishing the Asymptotic Size of Confidence
  Sets and Tests,} \emph{Journal of Econometrics}, 218, 496--531.

\bibitem[\protect\citeauthoryear{Andrews, Gill, Schank, and Upward}{Andrews
  et~al.}{2008}]{AndrewsGillSchankUpward08}
\textsc{Andrews, M.~J., L.~Gill, T.~Schank, and R.~Upward} (2008):
  \enquote{High Wage Workers and Low Wage Firms: Negative Assortative Matching
  or Limited Mobility Bias?} \emph{Journal of the Royal Statistical Society:
  Series A (Statistics in Society)}, 171, 673--697.

\bibitem[\protect\citeauthoryear{Arellano, Higgins, and Weidner}{Arellano
  et~al.}{2026}]{ArellanoHigginsWeidner26}
\textsc{Arellano, M., A.~Higgins, and M.~Weidner} (2026): \enquote{Quantile
  Methods in Large Panels with Fixed Effects,} arXiv:2610.38665.

\bibitem[\protect\citeauthoryear{Bonhomme, Holzheu, Lamadon, Manresa, Mogstad,
  and Setzler}{Bonhomme et~al.}{2023}]{BonhommeEtAl23}
\textsc{Bonhomme, S., K.~Holzheu, T.~Lamadon, E.~Manresa, M.~Mogstad, and
  B.~Setzler} (2023): \enquote{How Much Should We Trust Estimates of Firm
  Effects and Worker Sorting?} \emph{Journal of Labor Economics}, 41, 291--322.

\bibitem[\protect\citeauthoryear{Bonhomme, Lamadon, and Manresa}{Bonhomme
  et~al.}{2019}]{bonhomme2019distributional}
\textsc{Bonhomme, S., T.~Lamadon, and E.~Manresa} (2019): \enquote{A
  Distributional Framework for Matched Employer Employee Data,}
  \emph{Econometrica}, 87, 699--739.

\bibitem[\protect\citeauthoryear{Boyd and Vandenberghe}{Boyd and
  Vandenberghe}{2004}]{boydvandenberghe2004}
\textsc{Boyd, S. and L.~Vandenberghe} (2004): \emph{Convex Optimization},
  Cambridge: Cambridge University Press.

\bibitem[\protect\citeauthoryear{Card, Heining, and Kline}{Card
  et~al.}{2013}]{CardHeiningKline13}
\textsc{Card, D., J.~Heining, and P.~Kline} (2013): \enquote{Workplace
  Heterogeneity and the Rise of {W}est {G}erman Wage Inequality,}
  \emph{Quarterly Journal of Economics}, 128, 967--1015.

\bibitem[\protect\citeauthoryear{Chernozhukov, Fern\'andez-Val, and
  Weidner}{Chernozhukov et~al.}{2024}]{cfw}
\textsc{Chernozhukov, V., I.~Fern\'andez-Val, and M.~Weidner} (2024):
  \enquote{Network and Panel Quantile Effects via Distribution Regression,}
  \emph{Journal of Econometrics}, 240, 105009.

\bibitem[\protect\citeauthoryear{Chiang and Wu}{Chiang and Wu}{2026}]{chiangwu}\textsc{Chiang, H.~D. and W.~B. Wu} (2026): \enquote{Bias Correction and
  Inference for Panel Quantile Regression,} arXiv:2610.36443.

\bibitem[\protect\citeauthoryear{Dhaene and Jochmans}{Dhaene and Jochmans}{2015}]{DhaeneJochmans15}
\textsc{Dhaene, G. and K.~Jochmans} (2015): \enquote{Split-Panel Jackknife
  Estimation of Fixed-Effect Models,} \emph{Review of Economic Studies}, 82,
  991--1030.

\bibitem[\protect\citeauthoryear{Dvoretzky, Kiefer, and Wolfowitz}{Dvoretzky et~al.}{1956}]{dkw}
\textsc{Dvoretzky, A., J.~Kiefer, and J.~Wolfowitz} (1956): \enquote{Asymptotic
  Minimax Character of the Sample Distribution Function and of the Classical
  Multinomial Estimator,} \emph{Annals of Mathematical Statistics}, 27,
  642--669.

\bibitem[\protect\citeauthoryear{Fern\'andez-Val}{Fern\'andez-Val}{2005}]{FernandezVal05}
\textsc{Fern\'andez-Val, I.} (2005): \enquote{Bias Correction in Panel Data Models with Individual Specific Parameters,} Working Paper WP2005-041,
  Department of Economics, Boston University.

\bibitem[\protect\citeauthoryear{Fern\'andez-Val}{Fern\'andez-Val}{2009}]{FernandezVal09}
\textsc{Fern\'andez-Val, I.} (2009): \enquote{Fixed Effects Estimation of
  Structural Parameters and Marginal Effects in Panel Probit Models,}
  \emph{Journal of Econometrics}, 150, 71--85.

\bibitem[\protect\citeauthoryear{Fern\'andez-Val and Weidner}{Fern\'andez-Val and Weidner}{2016}]{FernandezValWeidner16}\textsc{Fern\'andez-Val, I. and M.~Weidner} (2016): \enquote{Individual and
  Time Effects in Nonlinear Panel Models with Large {$N$}, {$T$},} \emph{Journal of Econometrics}, 192, 291--312.

\bibitem[\protect\citeauthoryear{Galvao, Gu, and Volgushev}{Galvao
  et~al.}{2020}]{galvao}
\textsc{Galvao, A.~F., J.~Gu, and S.~Volgushev} (2020): \enquote{On the
  Unbiased Asymptotic Normality of Quantile Regression with Fixed Effects,}
  \emph{Journal of Econometrics}, 218, 178--215.

\bibitem[\protect\citeauthoryear{Galvao and Kato}{Galvao and
  Kato}{2016}]{galvaokato16}
\textsc{Galvao, A.~F. and K.~Kato} (2016): \enquote{Smoothed Quantile
  Regression for Panel Data,} \emph{Journal of Econometrics}, 193, 92--112.

\bibitem[\protect\citeauthoryear{Hahn and Newey}{Hahn and
  Newey}{2004}]{HahnNewey04}
\textsc{Hahn, J. and W.~Newey} (2004): \enquote{Jackknife and Analytical Bias Reduction for Nonlinear Panel Models,} \emph{Econometrica}, 72, 1295--1319.

\bibitem[\protect\citeauthoryear{Hall and Horowitz}{Hall and Horowitz}{1990}]{HallHorowitz90}
\textsc{Hall, P. and J. L. Horowitz} (1990): \enquote{Bandwidth Selection in Semiparametric Estimation of Censored Linear Regression Models,} \emph{Econometric Theory}, 6, 123--150.

\bibitem[\protect\citeauthoryear{Hjort and Pollard}{Hjort and Pollard}{1993}]{hjortpollard}
\textsc{Hjort, N.~L. and D.~Pollard} (1993): \enquote{Asymptotics for
  Minimisers of Convex Processes,} Unpublished manuscript; posted to arXiv in
  2011 as arXiv:1107.3806.

\bibitem[\protect\citeauthoryear{Hoeffding}{Hoeffding}{1963}]{hoeffding}
\textsc{Hoeffding, W.} (1963): \enquote{Probability Inequalities for Sums of
  Bounded Random Variables,} \emph{Journal of the American Statistical
  Association}, 58, 13--30.

\bibitem[\protect\citeauthoryear{Honor{\'e} and Kyriazidou}{Honor{\'e} and
  Kyriazidou}{2000}]{HonoreKyriazidou00}
\textsc{Honor{\'e}, B.~E. and E.~Kyriazidou} (2000): \enquote{Panel Data
  Discrete Choice Models with Lagged Dependent Variables,} \emph{Econometrica},
  68, 839--874.

\bibitem[\protect\citeauthoryear{Honor{\'e} and Weidner}{Honor{\'e} and
  Weidner}{2025}]{HonoreWeidner25}
\textsc{Honor{\'e}, B.~E. and M.~Weidner} (2025): \enquote{Moment Conditions
  for Dynamic Panel Logit Models with Fixed Effects,} \emph{Review of Economic
  Studies}, 92, 3112--3137.

\bibitem[\protect\citeauthoryear{Jochmans and Weidner}{Jochmans and
  Weidner}{2019}]{JochmansWeidner19}
\textsc{Jochmans, K. and M.~Weidner} (2019): \enquote{Fixed-Effect Regressions
  on Network Data,} \emph{Econometrica}, 87, 1543--1560.

\bibitem[\protect\citeauthoryear{Kato, Galvao, and Montes-Rojas}{Kato et~al.}{2012}]{kato}
\textsc{Kato, K., A.~F. Galvao, Jr., and G.~V. Montes-Rojas} (2012):
  \enquote{Asymptotics for Panel Quantile Regression Models with Individual
  Effects,} \emph{Journal of Econometrics}, 170, 76--91.

\bibitem[\protect\citeauthoryear{Kline, Saggio, and S{\o}lvsten}{Kline
  et~al.}{2020}]{KlineSaggioSolvsten20}
\textsc{Kline, P., R.~Saggio, and M.~S{\o}lvsten} (2020): \enquote{Leave-Out
  Estimation of Variance Components,} \emph{Econometrica}, 88, 1859--1898.

\bibitem[\protect\citeauthoryear{Knight}{Knight}{1998}]{knight}
\textsc{Knight, K.} (1998): \enquote{Limiting Distributions for {$L_1$}
  Regression Estimators under General Conditions,} \emph{Annals of Statistics},
  26, 755--770.

\bibitem[\protect\citeauthoryear{Koenker}{Koenker}{2004}]{koenker}
\textsc{Koenker, R.} (2004): \enquote{Quantile Regression for Longitudinal
  Data,} \emph{Journal of Multivariate Analysis}, 91, 74--89.

\bibitem[\protect\citeauthoryear{Koenker}{Koenker}{2005}]{koenkerbook}
---\hspace{-.1pt}---\hspace{-.1pt}--- (2005): \emph{Quantile Regression},
  Cambridge: Cambridge University Press.

\bibitem[\protect\citeauthoryear{Koenker}{Koenker}{2023}]{quantreg}
---\hspace{-.1pt}---\hspace{-.1pt}--- (2023): \emph{quantreg: Quantile
  Regression}, r package version 5.97.

\bibitem[\protect\citeauthoryear{Koenker and Bassett}{Koenker and Bassett}{1978}]{koenkerbassett}
\textsc{Koenker, R. and G.~Bassett, Jr.} (1978): \enquote{Regression
  Quantiles,} \emph{Econometrica}, 46, 33--50.

\bibitem[\protect\citeauthoryear{Koenker and Ng}{Koenker and
  Ng}{2003}]{koenkerng}
\textsc{Koenker, R. and P.~Ng} (2003): \enquote{{SparseM}: A Sparse Matrix
  Package for {R},} \emph{Journal of Statistical Software}, 8.

\bibitem[\protect\citeauthoryear{Ledoux and Talagrand}{Ledoux and
  Talagrand}{1991}]{ledouxtalagrand}
\textsc{Ledoux, M. and M.~Talagrand} (1991): \emph{Probability in {B}anach
  Spaces: Isoperimetry and Processes}, vol.~23 of \emph{Ergebnisse der
  Mathematik und ihrer Grenzgebiete (3)}, Berlin: Springer.

\bibitem[\protect\citeauthoryear{Massart}{Massart}{1990}]{massart}
\textsc{Massart, P.} (1990): \enquote{The Tight Constant in the
  {D}voretzky--{K}iefer--{W}olfowitz Inequality,} \emph{Annals of Probability},
  18, 1269--1283.

\bibitem[\protect\citeauthoryear{Portnoy and Koenker}{Portnoy and
  Koenker}{1997}]{portnoykoenker}
\textsc{Portnoy, S. and R.~Koenker} (1997): \enquote{The {G}aussian Hare and
  the {L}aplacian Tortoise: Computability of Squared-Error versus
  Absolute-Error Estimators,} \emph{Statistical Science}, 12, 279--300.

\bibitem[\protect\citeauthoryear{{R Core Team}}{{R Core Team}}{2024}]{rcore}
\textsc{{R Core Team}} (2024): \emph{R: A Language and Environment for
  Statistical Computing, Version 4.3.3}, R Foundation for Statistical
  Computing, Vienna, Austria.

\bibitem[\protect\citeauthoryear{van~der Vaart}{van~der Vaart}{1998}]{vdv}
\textsc{van~der Vaart, A.~W.} (1998): \emph{Asymptotic Statistics}, Cambridge
  Series in Statistical and Probabilistic Mathematics, Cambridge: Cambridge
  University Press.

\bibitem[\protect\citeauthoryear{van~der Vaart and Wellner}{van~der Vaart and
  Wellner}{2023}]{vdvw}
\textsc{van~der Vaart, A.~W. and J.~A. Wellner} (2023): \emph{Weak Convergence
  and Empirical Processes: With Applications to Statistics}, Springer Series in
  Statistics, Cham: Springer, second ed.

\bibitem[\protect\citeauthoryear{Verdier}{Verdier}{2020}]{Verdier20}
\textsc{Verdier, V.} (2020): \enquote{Estimation and Inference for Linear
  Models with Two-Way Fixed Effects and Sparsely Matched Data,} \emph{Review of
  Economics and Statistics}, 102, 1--16.

\end{thebibliography}
\end{document}